%% file: manuscript_final.tex
\documentclass[10pt,twocolumn,letterpaper]{article}

\usepackage{iccv}
\usepackage{times}

\usepackage{amsmath,amssymb,amsthm,amsbsy}
\usepackage{algorithm,algorithmic}
\usepackage{graphicx,subcaption}
\usepackage{booktabs,multirow}
\usepackage{enumitem}
\usepackage{cite}
\usepackage{appendix}

\theoremstyle{remark}
\newtheorem{remark}{Remark}

\newcommand{\expect}{\mathbb{E}}

\newcommand{\round}{\operatorname{round}}

\newcommand{\softplus}{\operatorname{softplus}}
\newcommand{\onehot}{\operatorname{onehot}}
\newcommand{\KL}{\operatorname{KL}}

\newcommand{\bx}{\mathbf{x}}

\newcommand{\bz}{\mathbf{z}}
\newcommand{\bu}{\mathbf{u}}
\newcommand{\bw}{\mathbf{w}}

\usepackage[breaklinks=true,bookmarks=false]{hyperref}

\iccvfinalcopy 

\def\iccvPaperID{3646} 

\ificcvfinal\fi

\begin{document}

\include{manuscript_final_main}
\include{manuscript_final_suppl}

\end{document}

%% file: manuscript_final_main.tex
\title{Variable Rate Deep Image Compression With a Conditional Autoencoder}

\author{Yoojin Choi, Mostafa El-Khamy, Jungwon Lee\\
SoC R\&D, Samsung Semiconductor Inc., San Diego, CA 92121, USA\\
{\tt\small \{yoojin.c,mostafa.e,jungwon2.lee\}@samsung.com}
}

\maketitle
\ificcvfinal\thispagestyle{empty}\fi

\begin{abstract}
In this paper, we propose a novel variable-rate learned image compression framework with a conditional autoencoder. Previous learning-based image compression methods mostly require training separate networks for different compression rates so they can yield compressed images of varying quality. In contrast, we train and deploy only one variable-rate image compression network implemented with a conditional autoencoder. We provide two rate control parameters, i.e., the Lagrange multiplier and the quantization bin size, which are given as conditioning variables to the network. Coarse rate adaptation to a target is performed by changing the Lagrange multiplier, while the rate can be further fine-tuned by adjusting the bin size used in quantizing the encoded representation. Our experimental results show that the proposed scheme provides a better rate-distortion trade-off than the traditional variable-rate image compression codecs such as JPEG2000 and BPG. Our model also shows comparable and sometimes better performance than the state-of-the-art learned image compression models that deploy multiple networks trained for varying rates.
\end{abstract}

\section{Introduction}

Image compression is an application of data compression for digital images to lower their storage and/or transmission requirements. Transform coding~\cite{goyal2001theoretical} has been successful to yield practical and efficient image compression algorithms such as JPEG~\cite{wallace1992jpeg} and JPEG2000~\cite{rabbani2002jpeg2000}. The transformation converts an input to a latent representation in the transform domain where lossy compression (that is typically a combination of quantization and lossless source coding) is more amenable and more efficient. For example, JPEG utilizes the discrete cosine transform (DCT) to convert an image into a sparse frequency domain representation. JPEG2000 replaces DCT with an enhanced discrete wavelet transform.

Deep learning is now leading many performance breakthroughs in various computer vision tasks~\cite{lecun2015deep}. Along with this revolutionary progress of deep learning, learned image compression also has derived significant interests~\cite{balle2017end,theis2017lossy,toderici2017full,rippel2017real,agustsson2017soft,mentzer2018conditional,balle2018variational,johnston2018improved,minnen2018joint,lee2019context}. In particular, non-linear transform coding designed with deep neural networks has advanced to outperform the classical image compression codecs sophisticatedly designed and optimized by domain experts, e.g., BPG~\cite{bellard2014bpg}, which is a still image version of the high efficiency video codec (HEVC) standard~\cite{sullivan2012overview}---we note that very recently, only a few of the learning-based image compression schemes have reached the performance of the state-of-the-art BPG codec on peak signal-to-noise ratio (PSNR), a metric based on mean squared error (MSE)~\cite{minnen2018joint,lee2019context}.

\begin{figure}[t!]
\centering
\includegraphics[width=0.85\columnwidth]{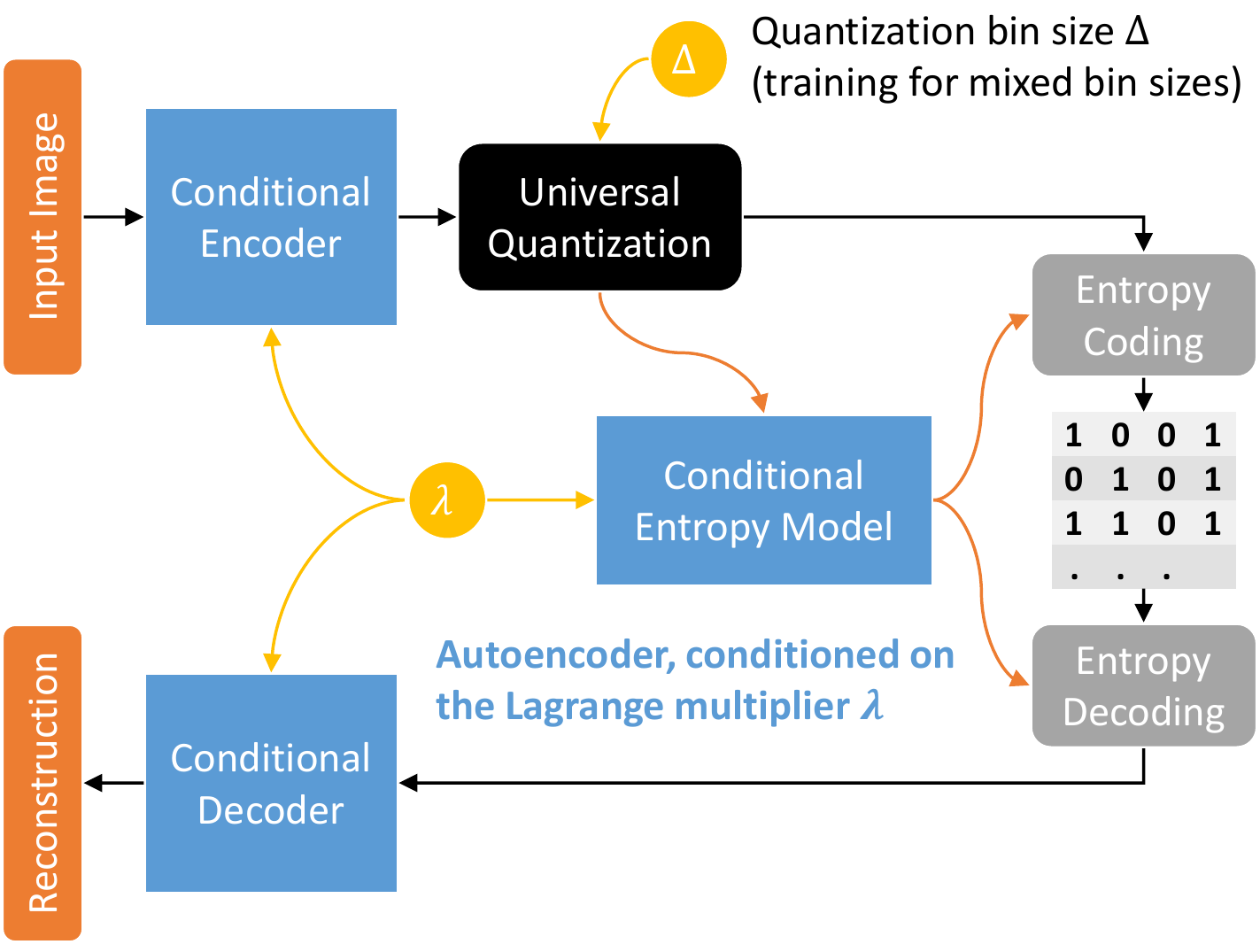}
\caption{Our variable-rate image compression model. We provide two knobs to vary the rate. First, we employ a conditional autoencoder, conditioned on the Lagrange multiplier~$\lambda$ that adapts the rate, and optimize the rate-distortion Lagrangian for various $\lambda$ values in one conditional model. Second, we train the model for mixed values of the quantization bin size~$\Delta$ so we can vary the rate by changing $\Delta$.\label{sec:fig:02}}
\vspace{-.7em}
\end{figure}

\setlength{\tabcolsep}{0.1em}
\begin{figure*}[t!]
\centering
{\scriptsize
\begin{tabular}{cccccc}
& Ground truth & Ours & BPG (4:4:4) & JPEG2000 & JPEG \\
\multirow{2}{*}[5.27em]{\includegraphics[width=0.3\textwidth]{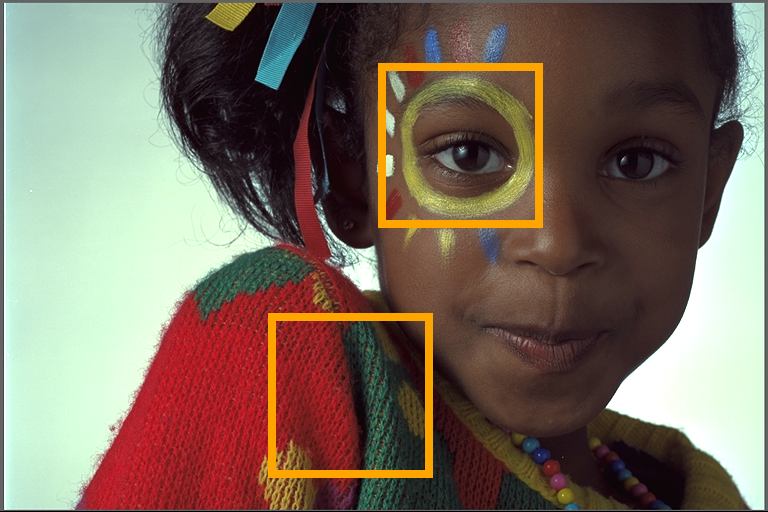}} &
\includegraphics[width=0.11\textwidth]{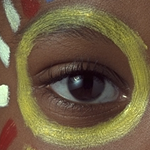} &
\includegraphics[width=0.11\textwidth]{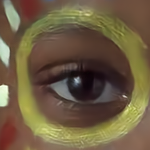} &
\includegraphics[width=0.11\textwidth]{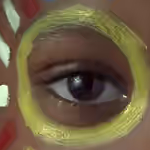} &
\includegraphics[width=0.11\textwidth]{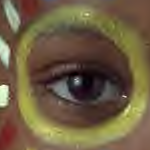}  &
\includegraphics[width=0.11\textwidth]{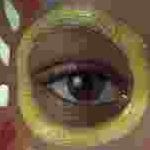} \\
&
\includegraphics[width=0.11\textwidth]{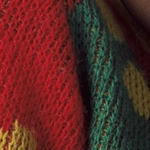} &
\includegraphics[width=0.11\textwidth]{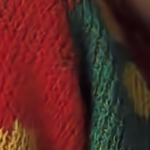} &
\includegraphics[width=0.11\textwidth]{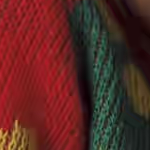} &
\includegraphics[width=0.11\textwidth]{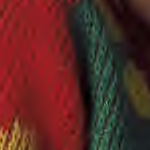}  &
\includegraphics[width=0.11\textwidth]{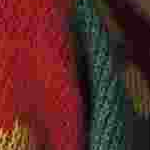} \\
\toprule
\multicolumn{2}{c}{Bits per pixel (BPP)} &  0.1697 &  0.1697 &  0.1702 &  0.1775 \\
\multicolumn{2}{c}{PSNR (dB)}            & 32.2332 & 31.9404 & 30.3140 & 27.3389 \\
\multicolumn{2}{c}{MS-SSIM}              &  0.9602 &  0.9539 &  0.9369 &  0.8669 \\
\bottomrule
\end{tabular}
}
\caption{PSNR and MS-SSIM comparison of our model and classical image compression algorithms (BPG, JPEG2000, and JPEG). We adapt the rate by changing the Lagrange multiplier~$\lambda$ and the quantization bin size~$\Delta$ to match the rate of BPG. In this example, we observe $0.3$ dB PSNR gain over the state-of-the-art BPG codec. A perceptual measure, MS-SSIM, also improves. Visually, our method provides better quality with less artifacts than the classical image compression codecs.\label{sec:train:fig:01}}
\vspace{-.7em}
\end{figure*}



The resemblance of non-linear transform coding and autoencoders has been established and exploited for image compression in \cite{balle2017end,theis2017lossy}---an encoder transforms an image (a set of pixels) into a latent representation in a lower dimensional space, and a decoder performs an approximate inverse transform that converts the latent representation back to the image. The transformation is desired to yield a latent representation with the smallest entropy, given a distortion level, since the entropy is the minimum rate achievable with lossless entropy source coding~\cite[Section~5.3]{cover2012elements}. In practice, however, it is generally not straightforward to calculate and optimize the exact entropy of a latent representation. Hence, the rate-distortion (R-D) trade-off is optimized by minimizing an entropy estimate of a latent representation provided by an autoencoder at a target quality. To improve compression efficiency, recent methods have focused on developing accurate entropy estimation models~\cite{agustsson2017soft,mentzer2018conditional,balle2018variational,minnen2018joint,lee2019context} with sophisticated density estimation techniques such as variational Bayes and autoregressive context modeling.

Given a model that provides an accurate entropy estimate of a latent representation, the previous autoencoder-based image compression frameworks optimize their networks by minimizing the weighted sum of the R-D pairs using the method of Lagrange multipliers. The Lagrange multiplier~$\lambda$ introduced in the Lagrangian (see \eqref{sec:prelim:eq:02}) is treated as a hyper-parameter to train a network for a desired trade-off between the rate and the quality of compressed images. This implies that one needs to train and deploy separate networks for rate adaptation. One way is to re-train a network while varying the Lagrange multiplier. However, this is impractical when we operate at a broad range of the R-D curve with fine resolution and the size of each network is large.

In this paper, we suggest training and deploying only one variable-rate image compression network that is capable of rate adaptation. In particular, we propose a conditional autoencoder, conditioned on the Lagrange multiplier, i.e., the network takes the Lagrange multiplier as an input and produces a latent representation whose rate depends on the input value. Moreover, we propose training the network with mixed quantization bin sizes, which allows us to adapt the rate by adjusting the bin size applied to the quantization of a latent representation. Coarse rate adaptation to a target is achieved by varying the Lagrange multiplier in the conditional model, while fine rate adaptation is done by tuning the quantization bin size. We illustrate our variable-rate image compression model in Figure~\ref{sec:fig:02}.

Conditional autoencoders have been used for conditional generation~\cite{sohn2015learning,van2016conditional}, where their conditioning variables are typically labels, attributes, or partial observations of the target output. However, our conditional autoencoder takes a hyper-parameter, i.e., the Lagrange multiplier, of the optimization problem as its conditioning variable. We basically solve multiple objectives using one conditional network, instead of solving them individually using separate non-conditional networks (each optimized for one objective), which is new to the best of our knowledge.

We also note that variable-rate models using recurrent neural networks (RNNs) were proposed in \cite{toderici2017full,johnston2018improved}. However, the RNN-based models require progressive encoding and decoding, depending on the target image quality. The increasing number of iterations to obtain a higher-quality image is not desirable in certain applications and platforms. Our variable-rate model is different from the RNN-based models. Our model is based on a conditional autoencoder that needs no multiple iterations, while the quality is controlled by its conditioning variables, i.e., the Lagrange multiplier and the quantization bin size. Our method also shows superior performance over the RNN-based models in \cite{toderici2017full,johnston2018improved}.

We evaluate the performance of our variable-rate image compression model on the Kodak image dataset~\cite{kodak1993kodak} for both the objective image quality metric, PSNR, and a perceptual score measured by the multi-scale structural similarity (MS-SSIM)~\cite{wang2003multiscale}. The experimental results show that our variable-rate model outperforms BPG in both PSNR and MS-SSIM metrics; an example from the Kodak dataset is shown in Figure~\ref{sec:train:fig:01}. 
Moreover, our model shows a comparable and sometime better R-D trade-off than the state-of-the-art learned image compression models~\cite{minnen2018joint,lee2019context} that outperform BPG by deploying multiple networks trained for different target rates.



\section{Preliminary} \label{sec:prelim}

We consider a typical autoencoder architecture consisting of encoder~$f_\phi(\bx)$ and decoder~$g_\theta(\bz)$, where $\bx$ is an input image and $\bz=\round_\Delta(f_\phi(\bx))$ is a quantized latent representation encoded from the input~$\bx$ with quantization bin size~$\Delta$; we let $\round_\Delta(x)=\Delta\round(x/\Delta)$, where $\round$ denotes element-wise rounding to the nearest integer. For now, we fix $\Delta=1$. Lossless entropy source coding, e.g., arithmetic coding~\cite[Section~13.3]{cover2012elements}, is used to generate a compressed bitstream from the quantized representation~$\bz$. Let $\expect_{p(x)}[A(x)]=\int A(x)p(x)dx$, where $p(x)$ is the probability density function of $x$. 

\textbf{Deterministic quantization}. Suppose that we take entropy source coding for the quantized latent variable~$\bz$ and achieve its entropy rate. The rate~$R$ and the squared L2 distortion~$D$ (i.e., the MSE loss) are given by
\begin{equation} \label{sec:prelim:eq:01}
\setlength{\abovedisplayskip}{.65em}
\setlength{\belowdisplayskip}{.65em}
\begin{split}
R_{\phi}&=\sum_{\bz}-P_\phi(\bz)\log_2P_\phi(\bz),\\
D_{\phi,\theta}&=\expect_{p(\bx)}[\|\bx-g_\theta(\round_\Delta(f_\phi(\bx)))\|_2^2],
\end{split}
\end{equation}
where $p(\bx)$ is the probability density function of all natural images, and $P_\phi(\bz)$ is the probability mass function of $\bz$ induced from encoder~$f_\phi(\bx)$ and $\round_\Delta$, which satisfies $P_\phi(\bz)=\int p(\bx)\delta(\bz-\round_\Delta(f_\phi(\bx)))d\bx$, where $\delta$ denotes the Dirac delta function. Using the method of Lagrange multipliers, the R-D optimization problem is given by
\begin{equation} \label{sec:prelim:eq:02}
\setlength{\abovedisplayskip}{.65em}
\setlength{\belowdisplayskip}{.65em}
\min_{\phi,\theta}\left\{D_{\phi,\theta}+\lambda R_\phi\right\},
\end{equation}
for $\lambda>0$; the scalar factor~$\lambda$ in the Lagrangian is called a Lagrange multiplier. The Lagrange multiplier is the factor that selects a specific R-D trade-off point (e.g., see \cite{ortega1998rate}).


\textbf{Relaxation with universal quantization}. The rate and the distortion provided in \eqref{sec:prelim:eq:01} are not differentiable for network parameter~$\phi$, due to $P_\phi(\bz)$ and $\round_\Delta$, and thus it is not straightforward to optimize \eqref{sec:prelim:eq:02} through gradient descent. It was proposed in \cite{balle2017end} to model the quantization error as additive uniform stochastic noise to relax the optimization of \eqref{sec:prelim:eq:02}. The same technique was adopted in \cite{balle2018variational,minnen2018joint,lee2019context}. In this paper, we instead propose employing universal quantization~\cite{ziv1985universal,zamir1992universal} to relax the problem (see Remark~\ref{sec:refined:univquant:remark:02}).

Universal quantization dithers every element of $f_\phi(\bx)$ with one common uniform random variable as follows:
\begin{equation} \label{sec:refined:univquant:eq:01}
\setlength{\abovedisplayskip}{.65em}
\setlength{\belowdisplayskip}{.65em}
\bz=\round_\Delta(f_\phi(\bx)+\bu)-\bu,
\ \ \ 
\bu=[U,U,\dots,U],
\end{equation}
where the dithering vector~$\mathbf{u}$ consists of repetitions of a single uniform random variable~$U$ with support~$[-\Delta/2,\Delta/2]$. We fix $\Delta=1$ just for now. In each dimension, universal quantization is effectively identical in distribution to adding uniform noise independent of the source, although the noise induced from universal quantization is dependent across dimensions. Note that universal quantization is approximated as a linear function of the unit slope (of gradient $1$) in the backpropagation of the network training. 

\begin{figure}[t!]
\centering
\includegraphics[width=.669\columnwidth]{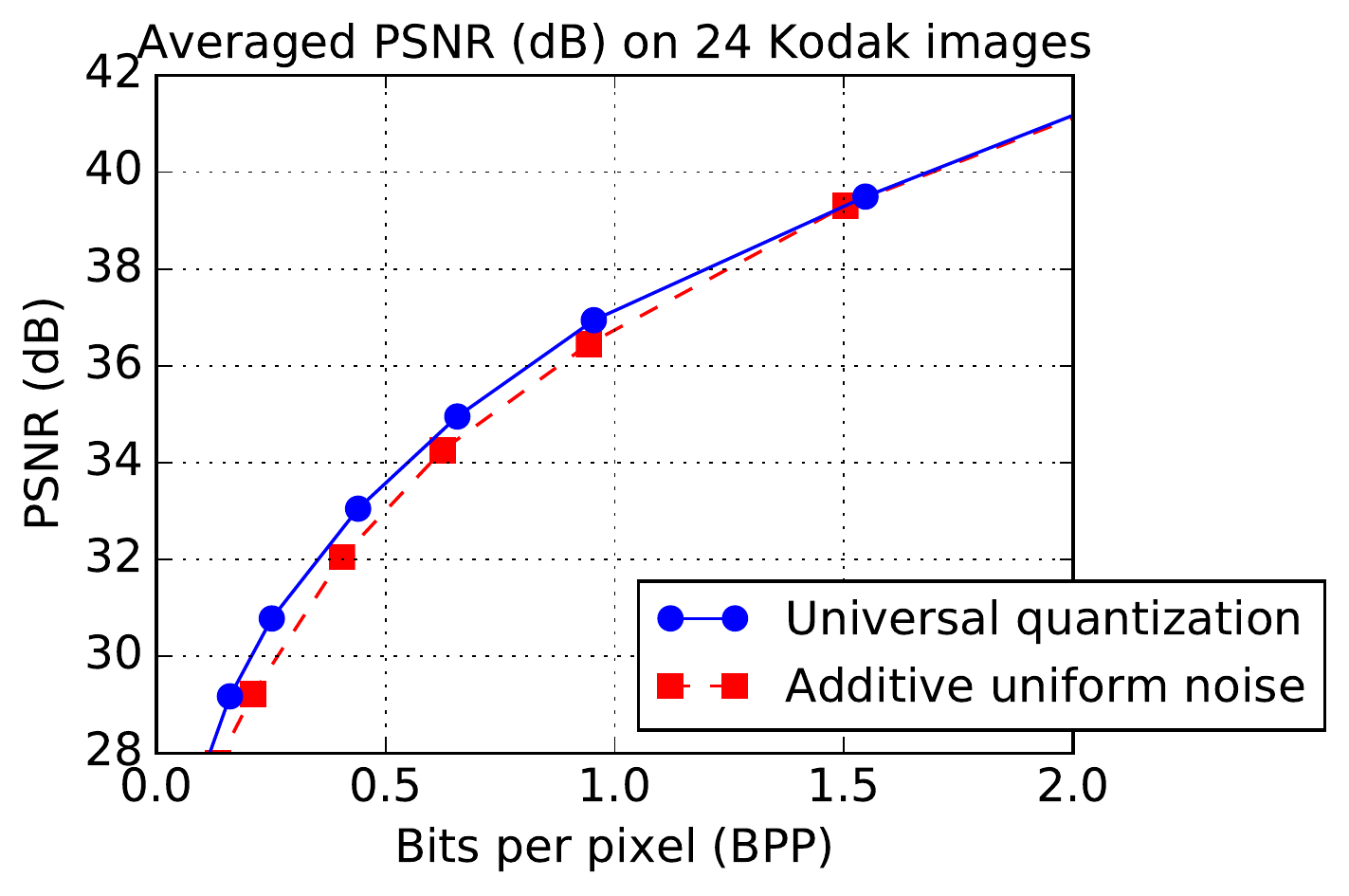}
\caption{The network trained with universal quantization gives higher PSNR than the one trained with additive uniform noise in our experiments on 24 Kodak images.
\label{sec:refined:univquant:fig:01}}
\vspace{-.7em}
\end{figure}

\begin{remark} \label{sec:refined:univquant:remark:01}
To our knowledge, we are the first to adopt universal quantization in the framework of training image compression networks. In \cite{choi2018universal}, universal quantization was used for efficient weight compression of deep neural networks, which is different from our usage here.
We observed from our experiments that our relaxation with universal quantization provides some gain over the conventional method of adding independent uniform noise (see Figure~\ref{sec:refined:univquant:fig:01}).
\end{remark}

\textbf{Differentiable R-D cost function}. Under the relaxation with universal quantization, similar to \eqref{sec:prelim:eq:01}, the rate and the distortion can be expressed as below:
\begin{equation} \label{sec:prelim:eq:04}
\setlength{\abovedisplayskip}{.65em}
\setlength{\belowdisplayskip}{.65em}
\begin{split}
R_{\phi}&=\expect_{p(\bx)p_\phi(\bz|\bx)}[-\log_2p_\phi(\bz)],\\
D_{\phi,\theta}&=\expect_{p(\mathbf{x})p_\phi(\mathbf{z}|\mathbf{x})}[\|\mathbf{x}-g_\theta(\bz)\|_2^2],
\end{split}
\end{equation}
where $p_\phi(\bz)=\int p(\bx)p_\phi(\bz|\bx)d\bx$. The stochastic quantization model makes $\bz$ have a continuous density~$p_\phi(\bz)$, which is a continuous relaxation of $P_\phi(\bz)$, but still $p_\phi(\bz)$ is usually intractable to compute. Thus, we further adopt approximation of $p_\phi(\bz)$ to a tractable density~$q_\theta(\bz)$ that is differentiable with respect to $\bz$ and $\theta$. Then, it follows that
\begin{equation} \label{sec:prelim:eq:05}
\setlength{\abovedisplayskip}{.65em}
\setlength{\belowdisplayskip}{.65em}
\begin{split}
R_{\phi}&=\expect_{p(\bx)p_\phi(\bz|\bx)}[-\log_2q_\theta(\bz)]-\KL(p_\phi(\bz)||q_\theta(\bz))\\
&\leq\expect_{p(\bx)p_\phi(\bz|\bx)}[-\log_2q_\theta(\bz)]\triangleq R_{\phi,\theta},
\end{split}
\end{equation}
where $\KL$ denotes Kullback-Leibler (KL) divergence~(e.g., see \cite[p.~19]{cover2012elements}); the equality in $\leq$ holds when $p_\phi(\bz)=q_\theta(\bz)$. The choice of $q_\theta(\bz)$ in our implementation is deferred to Section~\ref{sec:refined} (see \eqref{sec:refined:eq:01}--\eqref{sec:refined:eq:03}).

From \eqref{sec:prelim:eq:02} and \eqref{sec:prelim:eq:04}, approximating $R_{\phi}$ by its upperbound $R_{\phi,\theta}$ in \eqref{sec:prelim:eq:05}, the R-D optimization problem reduces to
\begin{equation} \label{sec:prelim:eq:06}
\setlength{\abovedisplayskip}{.65em}
\setlength{\belowdisplayskip}{.65em}
\min_{\phi,\theta}\expect_{p(\bx)p_\phi(\bz|\bx)}[\|\mathbf{x}-g_\theta(\bz)\|_2^2-\lambda\log_2q_\theta(\bz)],
\end{equation}
for $\lambda>0$. Optimizing a network for different values of $\lambda$, one can trade off the quality against the rate.

\begin{remark} \label{sec:refined:univquant:remark:02}
The objective function in \eqref{sec:prelim:eq:06} has the same form as auto-encoding variational Bayes~\cite{kingma2014auto}, given that the posterior~$p_\phi(\bz|\bx)$ is uniform.
This relation was already established in the previous works, and detailed discussions can be found in \cite{balle2017end,balle2018variational}. Our contribution in this section is to deploy universal quantization (see \eqref{sec:refined:univquant:eq:01}) to guarantee that the quantization error is uniform and independent of the source distribution, instead of artificially adding uniform noise, when generating random samples of $\bz$ from $p_\phi(\bz|\bx)$ in Monte Carlo estimation of \eqref{sec:prelim:eq:06}.
\end{remark}

\section{Variable rate image compression} \label{sec:var-rate}

To adapt the quality and the rate of compressed images, we basically need to optimize the R-D Lagrange function in \eqref{sec:prelim:eq:06} for varying values of the Lagrange multiplier~$\lambda$. That is, one has to train multiple networks or re-train a network while varying the Lagrange multiplier~$\lambda$. Training and deploying multiple networks are not practical, in particular when we want to cover a broad range of the R-D curve with fine resolution, and each network is of a large size. In this section, we develop a variable-rate model that can be deployed once and can be used to produce compressed images of varying quality with different rates, depending on user's requirements, with no need of re-training. 

\subsection{Conditional autoencoder} \label{sec:var-rate:condconv}

\begin{figure}[t!]
\centering
\includegraphics[width=.7\columnwidth]{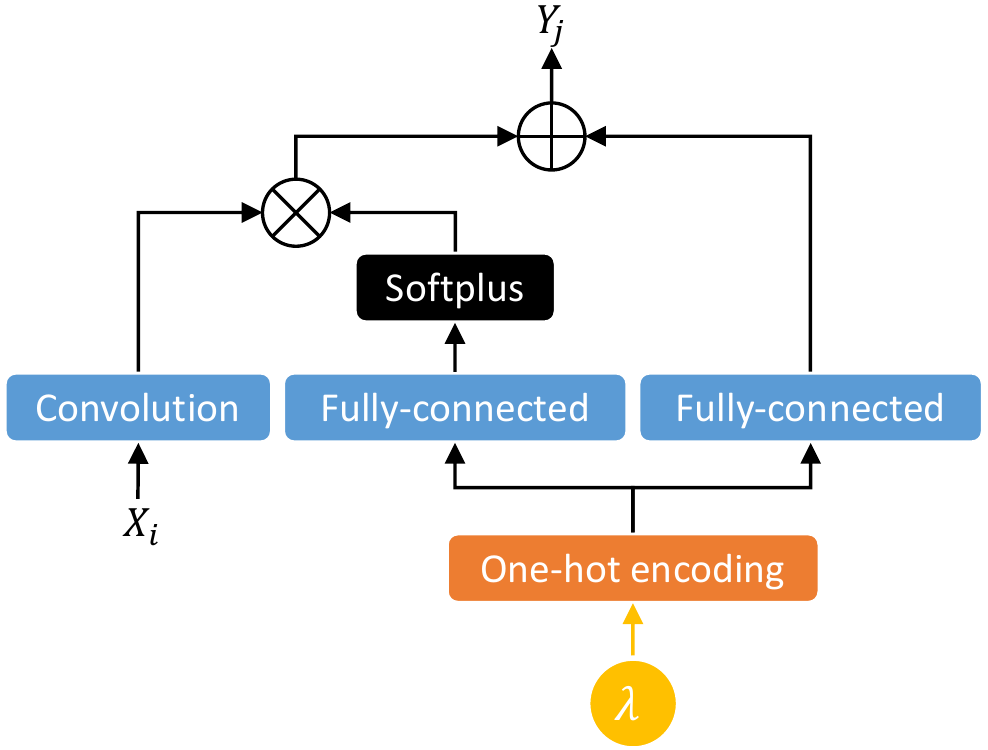}
\caption{Conditional convolution, conditioned on the Lagrange multiplier~$\lambda$, which produces a different output depending on the input Lagrange multiplier~$\lambda$.
\label{sec:var-rate:condconv:fig:01}}
\vspace{-.7em}
\end{figure}

To avoid training and deploying multiple networks, we propose training one conditional autoencoder, conditioned on the Lagrange multiplier~$\lambda$. The network takes $\lambda$ as a conditioning input parameter, along with the input image, and produces a compressed image with varying rate and distortion depending on the conditioning value of $\lambda$. To this end, the rate and distortion terms in \eqref{sec:prelim:eq:04} and \eqref{sec:prelim:eq:05} are altered into
\[
\setlength{\abovedisplayskip}{.65em}
\setlength{\belowdisplayskip}{.65em}
\begin{split}
R_{\phi,\theta}(\lambda)&=\expect_{p(\bx)p_\phi(\bz|\bx,\lambda)}[-\log_2q_\theta(\bz|\lambda)],\\
D_{\phi,\theta}(\lambda)&=\expect_{p(\bx)p_\phi(\mathbf{z}|\mathbf{x},\lambda)}[\|\mathbf{x}-g_\theta(\bz,\lambda)\|_2^2],
\end{split}
\]
for $\lambda\in\Lambda$, where $\Lambda$ is a pre-defined finite set of Lagrange multiplier values, and then we minimize the following combined objective function:
\begin{equation} \label{sec:var-rate:condconv:eq:01}
\setlength{\abovedisplayskip}{.65em}
\setlength{\belowdisplayskip}{.65em}
\min_{\phi,\theta}\sum_{\lambda\in\Lambda}\left(D_{\phi,\theta}(\lambda)+\lambda R_{\phi,\theta}(\lambda)\right).
\end{equation}

To implement a conditional autoencoder, we develop the conditional convolution, conditioned on the Lagrange multiplier~$\lambda$, as shown in Figure~\ref{sec:var-rate:condconv:fig:01}. Let $X_i$ be a 2-dimensional (2-D) input feature map of channel~$i$ and $Y_j$ be a 2-D output feature map of channel~$j$. Let $W_{i,j}$ be a 2-D convolutional kernel for input channel~$i$ and output channel~$j$. Our conditional convolution yields
\begin{equation} \label{sec:var-rate:condconv:eq:02}
\setlength{\abovedisplayskip}{.65em}
\setlength{\belowdisplayskip}{.65em}
Y_j=s_j(\lambda)\sum_{i}X_i*W_{i,j}+b_j(\lambda),
\end{equation}
where $*$ denotes 2-D convolution. The channel-wise scaling factor and the additive bias term depend on $\lambda$ by
\begin{equation} \label{sec:var-rate:condconv:eq:03}
\setlength{\abovedisplayskip}{.65em}
\setlength{\belowdisplayskip}{.65em}
\begin{split}
s_j(\lambda)&=\softplus(u_j^T\onehot_\Lambda(\lambda)),\\
b_j(\lambda)&=v_j^T\onehot_\Lambda(\lambda),
\end{split}
\end{equation}
where $u_j$ and $v_j$ are the fully-connected layer weight vectors of length $|\Lambda|$ for output channel~$j$; $T$ denotes the transpose, $\softplus(x)=\log(1+e^x)$, and $\onehot_\Lambda(\lambda)$ is one-hot encoding of $\lambda$ over $\Lambda$.


\begin{remark}
The proposed conditional convolution is similar to the one proposed by conditional PixelCNN~\cite{van2016conditional}. In \cite{van2016conditional}, conditioning variables are typically labels, attributes, or partial observations of the target output, while our conditioning variable is the Lagrange multiplier, which is the hyper-parameter that trades off the quality against the rate in the compression problem. A gated-convolution structure is presented in \cite{van2016conditional}, but we develop a simpler structure so that the additional computational cost of conditioning is marginal.
\end{remark}

\subsection{Training with mixed bin sizes} \label{sec:var-rate:mixedbin}

\setlength{\tabcolsep}{0.1em}
\begin{figure*}[t!]
\centering
{\small
\begin{tabular}{ccc}
\includegraphics[height=.215\textwidth]{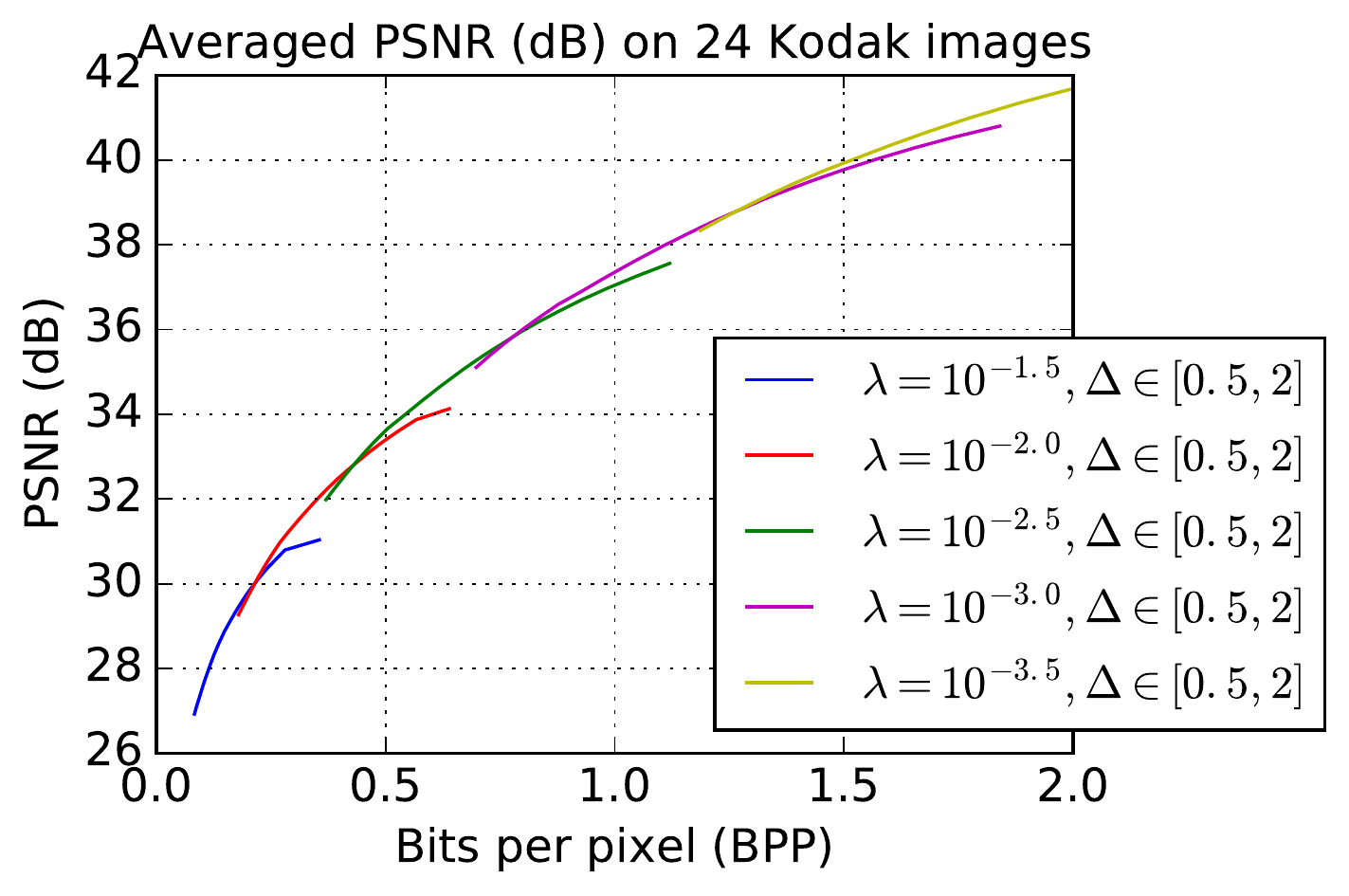} &
\includegraphics[height=.215\textwidth]{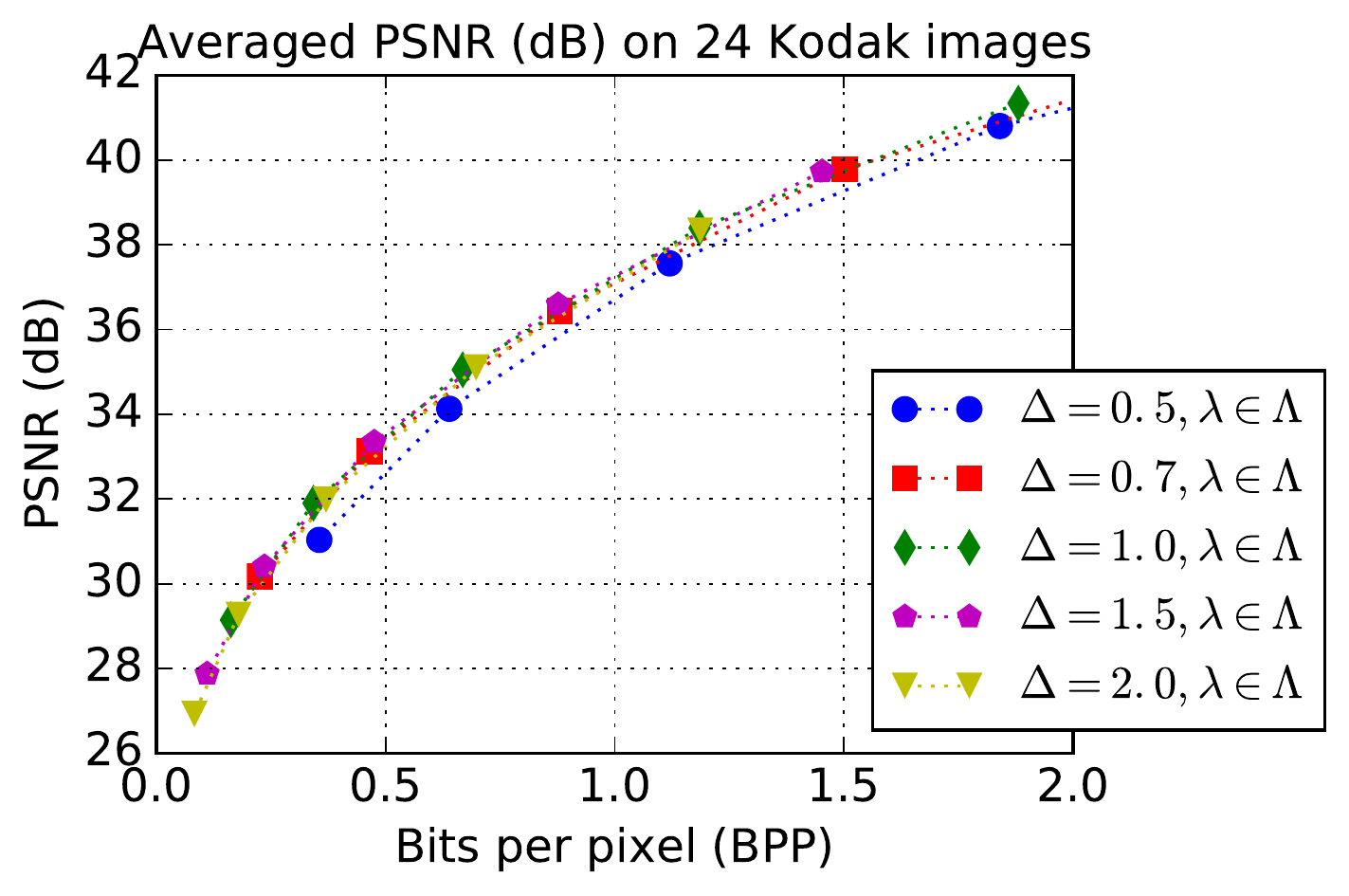} &
\includegraphics[height=.215\textwidth]{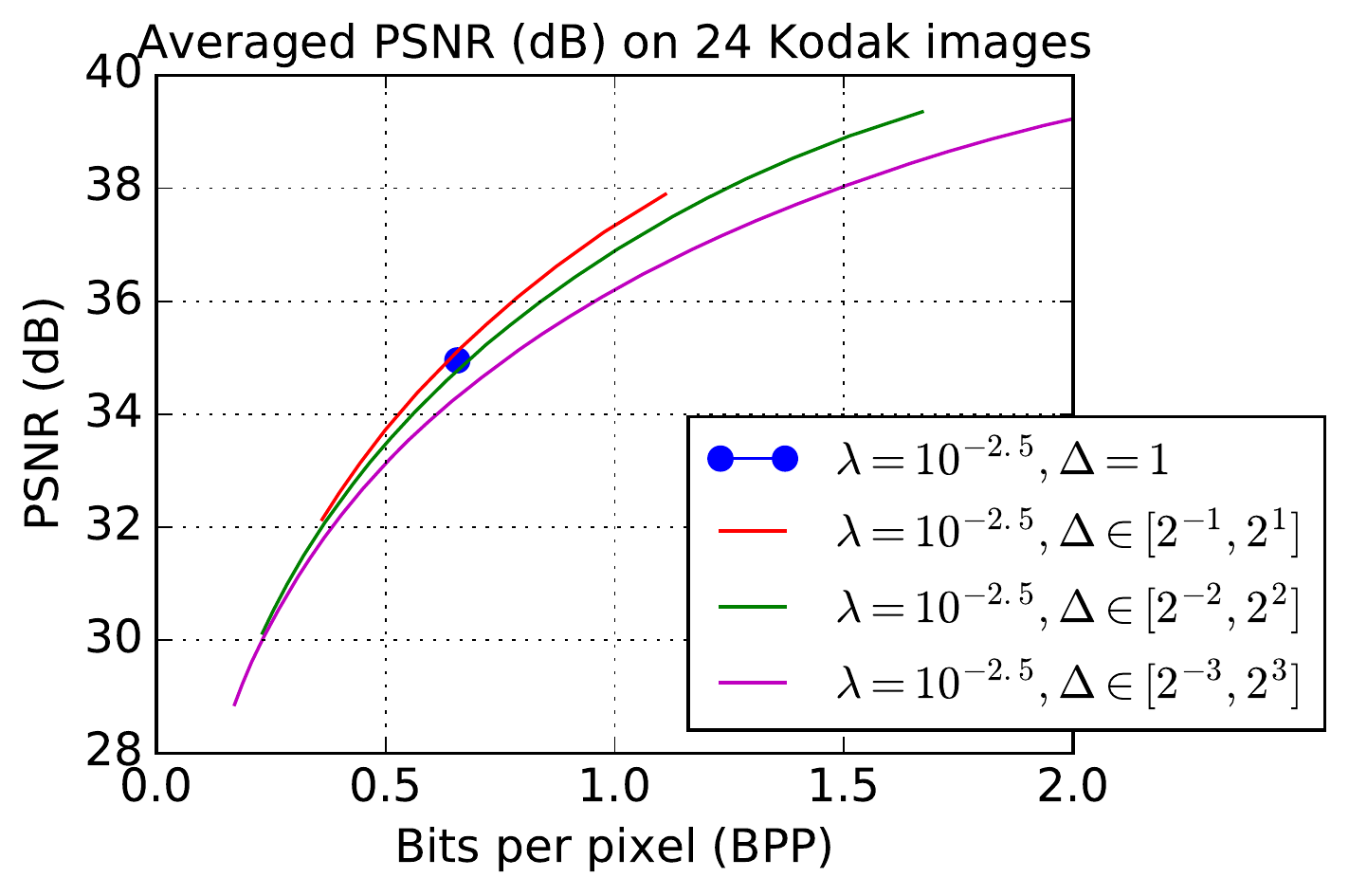} \\
(a) Vary $\Delta\in[0.5,2]$ for fixed $\lambda\in\Lambda$ & (b) Vary $\lambda\in\Lambda$ for fixed $\Delta$ & (c) Vary the mixing range of $\Delta$ in training\\
\end{tabular}
}
\vspace{-.5em}
\caption{In (a,b), we show how we can adapt the rate in our variable-rate model by changing the Lagrange multiplier~$\lambda$ and the quantization bin size~$\Delta$. In (a), we vary $\Delta$ within $[0.5,2]$ for each fixed $\lambda\in\Lambda$ in \eqref{sec:exp:setup:eq:01}. In (b), we change $\lambda$ in $\Lambda$ while fixing $\Delta$ for some selected values.
In (c), we compare PSNR when models are trained for mixed bin sizes of different ranges.
\label{sec:var-rate:mixedbin:fig:01}}
\vspace{-2em}
\end{figure*}

We established a variable-rate conditional autoencoder model, conditioned on the Lagrange multiplier~$\lambda$ in the previous subsection, but only finite discrete points in the R-D curve can be obtained from it, since $\lambda$ is selected from a pre-determined finite set~$\Lambda$.\footnote{
The conditioning part can be modified to take continuous $\lambda$ values,
which however did not produce good results in our trials. 
}
To extend the coverage to the whole continuous range of the R-D curve, we develop another (continuous) knob to control the rate, i.e., the quantization bin size.

Recall that in the previous R-D formulation~\eqref{sec:prelim:eq:01}, we fixed the quantization bin size~$\Delta=1$, i.e., we simply used $\round$ for quantization. In actual inference, we can change the bin size to adapt the rate---the larger the bin size, the lower the rate. However, the performance naturally suffers from mismatched bin sizes in training and inference. For a trained network to be robust and accurate for varying bin sizes, we propose training (or fine-tuning) it with mixed bin sizes.

In training, we draw a uniform noise in \eqref{sec:refined:univquant:eq:01} for various noise levels, i.e., for random $\Delta$. The range of $\Delta$ and the mixing distribution within the range are design choices. In our experiments, we choose $\Delta=2^b$, where $b$ is uniformly drawn from $[-1,1]$ so we can cover $\Delta\in[0.5,2]$. The larger the range of $b$, we optimize a network for a broader range of the R-D curve, but the performance also degrades. In Figure~\ref{sec:var-rate:mixedbin:fig:01}(c), we compare the R-D curves obtained from the networks trained with mixed bin sizes of different ranges; we used fixed $\lambda=10^{-2.5}$ in training the networks just for this experiment.
We found that mixing bin sizes in $\Delta\in[0.5,2]$ yields the best performance, although the coverage is limited, which is not a problem since we can cover large-scale rate adaptation by changing the input Lagrange multiplier in our conditional model (see Figure~\ref{sec:var-rate:mixedbin:fig:01}~(a,b)).

In summary, we solve the following optimization:
\begin{equation} \label{sec:var-rate:mixedbin:eq:01}
\setlength{\abovedisplayskip}{.8em}
\setlength{\belowdisplayskip}{.8em}
\min_{\phi,\theta}\sum_{\lambda\in\Lambda}\expect_{p(\Delta)}[D_{\phi,\theta}(\lambda,\Delta)+\lambda R_{\phi,\theta}(\lambda,\Delta)],
\end{equation}
where $p(\Delta)$ is a pre-defined mixing density for $\Delta$, and
\begin{equation} \label{sec:var-rate:mixedbin:eq:02}
\setlength{\abovedisplayskip}{.8em}
\setlength{\belowdisplayskip}{.8em}
\begin{split}
R_{\phi,\theta}(\lambda,\Delta)&=\expect_{p(\bx)p_\phi(\bz|\bx,\lambda,\Delta)}[-\log_2q_\theta(\bz|\lambda,\Delta)],\\
D_{\phi,\theta}(\lambda,\Delta)&=\expect_{p(\bx)p_\phi(\mathbf{z}|\mathbf{x},\lambda,\Delta)}[\|\mathbf{x}-g_\theta(\bz,\lambda)\|_2^2].
\end{split}
\end{equation}

\begin{remark} \label{sec:var-rate:mixedbin:remark:01}
In training, we compute neither the summation over $\lambda\in\Lambda$ nor the expectation over $p(\Delta)$ in \eqref{sec:var-rate:mixedbin:eq:01}. Instead, we randomly select $\lambda$ uniformly from $\Lambda$ and draw $\Delta$ from $p(\Delta)$ for each image to compute its individual R-D cost, and then we use the average R-D cost per batch as the loss for gradient descent, which makes the training scalable.
\end{remark}

\subsection{Inference}

\textbf{Rate adaptation}. The rate increases, as we decrease the Lagrange multiplier~$\lambda$ and/or the quantization bin size~$\Delta$. In Figure~\ref{sec:var-rate:mixedbin:fig:01}(a,b), we show how the rate varies as we change $\lambda$ and $\Delta$. In (a), we change $\Delta$ within $[0.5,2]$ for each fixed $\lambda\in\Lambda$ from \eqref{sec:exp:setup:eq:01}. In (b), we vary $\lambda$ in $\Lambda$ while fixing $\Delta$ for some selected values. Given a user's target rate, large-scale discrete rate adaptation is achieved by changing $\lambda$, while fine continuous rate adaptation can be performed by adjusting $\Delta$ for fixed $\lambda$. When the R-D curves overlap at the target rate (e.g., see $0.5$ BPP in Figure~\ref{sec:var-rate:mixedbin:fig:01}(a)), we select the combination of $\lambda$ and $\Delta$ that produces better performance.\footnote{In practice, one can make a set of pre-selected combinations of $\lambda$ and $\Delta$, similar to the set of quality factors in JPEG or BPG.}

\textbf{Compression}. After selecting $\lambda\in\Lambda$, we do one-hot encoding of $\lambda$ and use it in all conditional convolutional layers to encode a latent representation of the input. Then, we perform regular deterministic quantization on the encoded representation with the selected quantization bin size~$\Delta$.
The quantized latent representation is then finally encoded into a compressed bitstream with entropy coding, e.g., arithmetic coding; we additionally need to store the values of the conditioning variables, $\lambda$ and $\Delta$, used in encoding.

\textbf{Decompression}. We decode the compressed bitstream. We also retrieve $\lambda$ and $\Delta$ used in encoding from the compressed bitstream. We restore the quantized latent representation from the decoded integer values by multiplying them with the quantization bin size~$\Delta$. The restored latent representation is then fed to the decoder to reconstruct the image. The value of $\lambda$ used in encoding is again used in all deconvolutional layers, for conditional generation.

\section{Refined probabilistic model} \label{sec:refined}

\begin{figure}[t!]
\centering
\includegraphics[width=0.746\columnwidth]{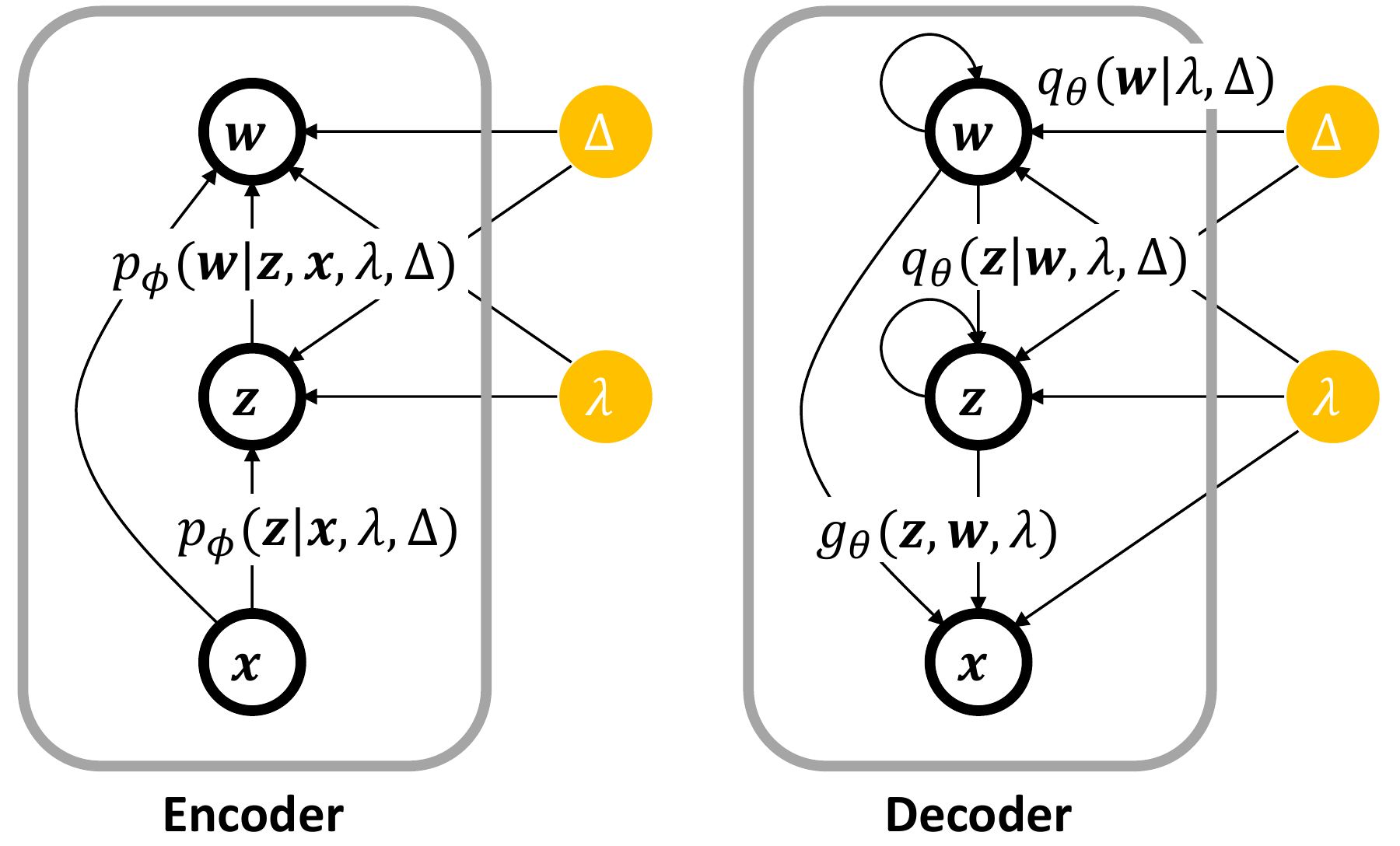}
\caption{A graph representation of our refined variable-rate image compression model.
\label{sec:refined:entropy:fig:01}}
\vspace{-.7em}
\end{figure}

In this section, we discuss how we refine the baseline model in the previous section to improve the performance. The model refinement is orthogonal to the rate adaptation schemes in Section~\ref{sec:var-rate}.
From \eqref{sec:var-rate:mixedbin:eq:02}, we introduce a secondary latent variable $\bw$ that depends on $\bx$ and $\bz$ to yield
{
\setlength{\abovedisplayskip}{.5em}
\setlength{\belowdisplayskip}{.5em}
\begin{multline*}
R_{\phi,\theta}(\lambda,\Delta)
=\expect_{p(\bx)p_\phi(\bz|\bx,\lambda,\Delta)p_\phi(\bw|\bz,\bx,\lambda,\Delta)} \\
[-\log_2(q_\theta(\bw|\lambda,\Delta)q_\theta(\bz|\bw,\lambda,\Delta))],
\end{multline*}
\vspace{-2em}
\begin{multline*}
D_{\phi,\theta}(\lambda,\Delta)
=\expect_{p(\bx)p_\phi(\bz|\bx,\lambda,\Delta)p_\phi(\bw|\bz,\bx,\lambda,\Delta)}  \\
[\|\mathbf{x}-g_\theta(\bz,\bw,\lambda)\|_2^2].
\end{multline*}
For compression, we encode $\bz$ from $\bx$, and then we further encode $\bw$ from $\bz,\bx$.} The encoded representations~$\bz,\bw$ are entropy-coded based on $q_\theta(\bw|\lambda,\Delta),q_\theta(\bz|\bw,\lambda,\Delta)$, respectively. For decompression, given $q_\theta(\bw|\lambda,\Delta)$, we decode $\bw$, which is then used to compute $q_\theta(\bz|\bw,\lambda,\Delta)$ and to decode $\bz$. This model is further refined by introducing autoregressive models for $q_\theta(\bw|\lambda,\Delta)$ and $q_\theta(\bz|\bw,\lambda,\Delta)$ as below:
\begin{equation} \label{sec:refined:eq:01}
\begin{split}
q_\theta(\bw|\lambda,\Delta)&=\prod\nolimits_{i}q_\theta(w_i|w_{<i},\lambda,\Delta), \\
q_\theta(\bz|\bw,\lambda,\Delta)&=\prod\nolimits_{i}q_\theta(z_i|z_{<i},\bw,\lambda,\Delta),
\end{split}
\end{equation}
where $a_i$ is the $i$-th element of $\mathbf{a}$, and $a_{<i}=[a_1,\dots,a_{i-1}]$. In Figure~\ref{sec:refined:entropy:fig:01}, we illustrate a graph representation of our refined variable-rate image compression model.

In our experiments, we use
\begin{equation} \label{sec:refined:eq:02}
\setlength{\abovedisplayskip}{.5em}
q_\theta(z_i|z_{<i},\bw,\lambda,\Delta)
=\int_{z_i-\Delta/2}^{z_i+\Delta/2}\frac{1}{\Delta\sigma_i}f_{\mathcal{N}}\left(\frac{x-\mu_i}{\sigma_i}\right)dx,
\end{equation}
where $\mu_i=\mu_\theta(z_{<i},\bw,\lambda)$, $\sigma_i^2=\sigma_\theta^2(z_{<i},\bw,\lambda)$, and $f_{\mathcal{N}}$ denotes the standard normal density; 
$\mu_\theta$ and $\sigma_\theta^2$ are parameterized with autoregressive neural networks, e.g., consisting of masked convolutions~\cite{van2016conditional}, which are also conditioned on $\lambda$ as in Figure~\ref{sec:var-rate:condconv:fig:01}.
Similarly, we let
\begin{equation} \label{sec:refined:eq:03}
\setlength{\abovedisplayskip}{.5em}
q_\theta(w_i|w_{<i},\lambda,\Delta)
=\int_{w_i-\Delta/2}^{w_i+\Delta/2}\frac{1}{\Delta\zeta_i}f_\psi\left(\frac{x-\nu_i}{\zeta_i}\right)dx,
\end{equation}
where $\nu_i=\nu_\theta(w_{<i},\lambda)$, $\zeta_i^2=\zeta_\theta^2(w_{<i},\lambda)$, and $f_\psi$ is designed as a univariate density model parameterized with a neural network as described in \cite[Appendix~6.1]{balle2018variational}. 

\begin{remark}
Setting aside the conditioning parts, the refined model can be viewed as a hierarchical autoencoder (e.g., see \cite{tomczak2018vae}). It is also similar to the one in \cite{minnen2018joint} with the differences summarized in Appendix~\ref{app:02}.
\end{remark}

\section{Experiments} \label{sec:exp}

\begin{figure*}
\centering
\includegraphics[width=.967\textwidth]{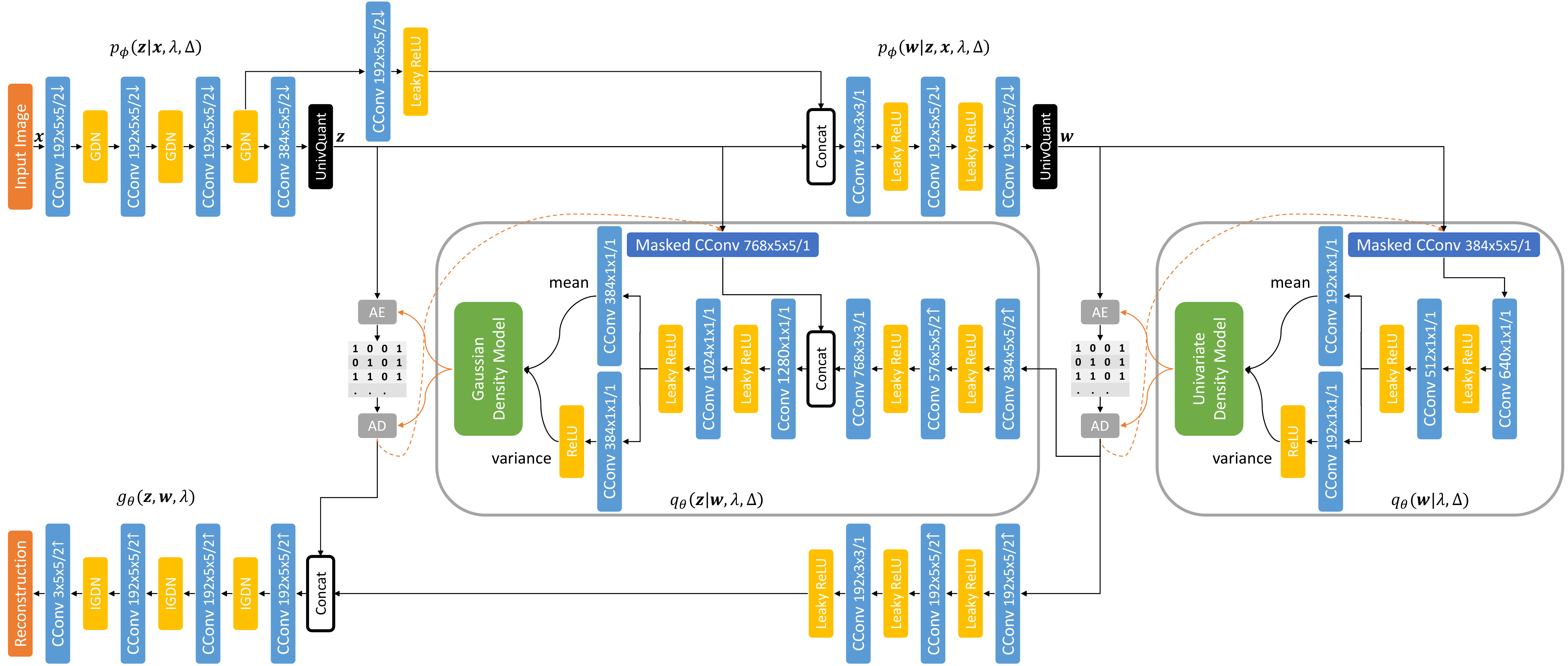}
\caption{\textbf{UnivQuant} denotes universal quantization with the quantization bin size~$\Delta$. \textbf{AE} and \textbf{AD} are arithmetic encoding and decoding, respectively. \textbf{Concat} implies concatenation. \textbf{GDN} stands for generalized divisive normalization, and \textbf{IGDN} is inverse GDN~\cite{balle2015density}. The convolution parameters are denoted as $\text{\# filters}\times\text{kernel height}\times\text{kernel width}~/~\text{stride}$, where $\uparrow$ and $\downarrow$ indicate upsampling and downsampling, respectively. \textbf{CConv} denotes conditional convolution, conditioned on the Lagrange multiplier~$\lambda$ (see Figure~\ref{sec:var-rate:condconv:fig:01}). All convolution and masked convolution blocks employ conditional convolutions. Upsampling convolutions are implemented as the deconvolution. Masked convolutions are implemented as in \cite{van2016conditional}.\label{sec:exp:setup:fig:01}}
\vspace{-.7em}
\end{figure*}


We illustrate the network architecture that we used in our experiments in Figure~\ref{sec:exp:setup:fig:01}. We emphasize that all convolution (including masked convolution) blocks employ conditional convolutions (see Figure~\ref{sec:var-rate:condconv:fig:01} in Section~\ref{sec:var-rate:condconv}).

\textbf{Training}. For a training dataset, we used the ImageNet ILSVRC 2012 dataset~\cite{russakovsky2015imagenet}. We resized the training images so that the shorter of the width and the height is $256$, and we extracted $256\times256$ patches at random locations. In addition to the ImageNet dataset, we used the training dataset provided in the Workshop and Challenge on Learned Image Compression (CLIC)\footnote{https://www.compression.cc}. For the CLIC training dataset, we extracted $256\times256$ patches at random locations without resizing. We used Adam optimizer~\cite{kingma2014adam} and trained a model for $50$ epochs, where each epoch consists of $40$k batches and the batch size is set to $8$. The learning rate was set to be $10^{-4}$ initially, and we decreased the learning rate to $10^{-5}$ and $10^{-6}$ at $20$ and $40$ epochs, respectively.

\begin{figure*}[t!]
\centering
\includegraphics[height=.224\textwidth]{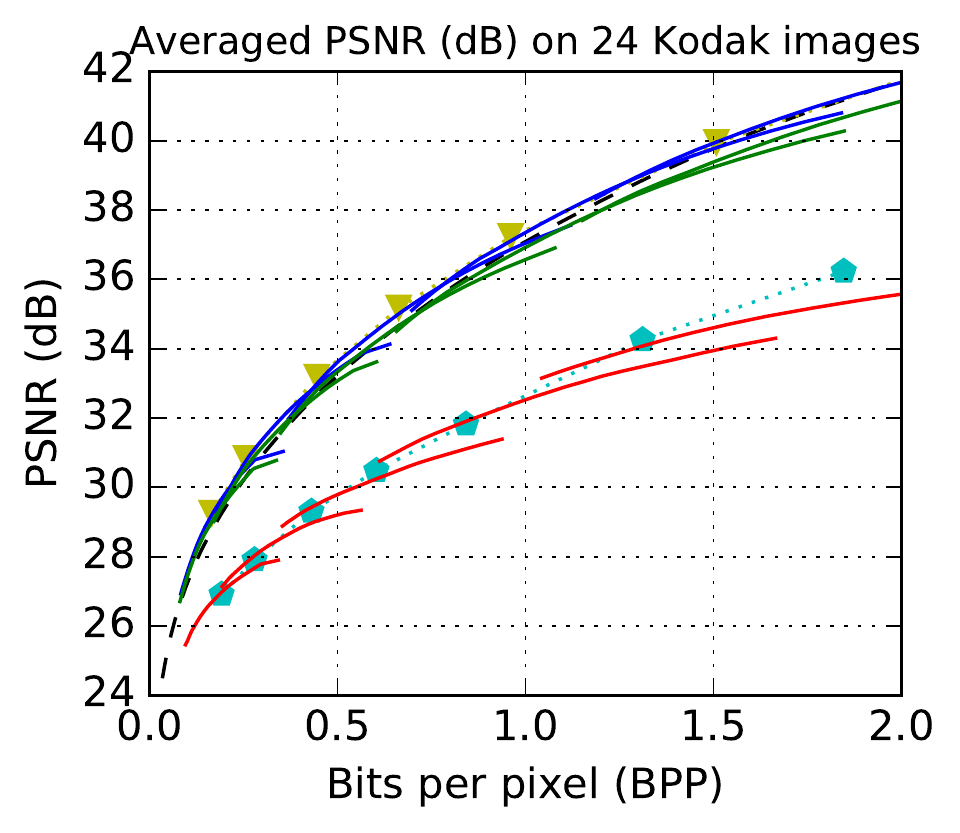}
\includegraphics[height=.224\textwidth]{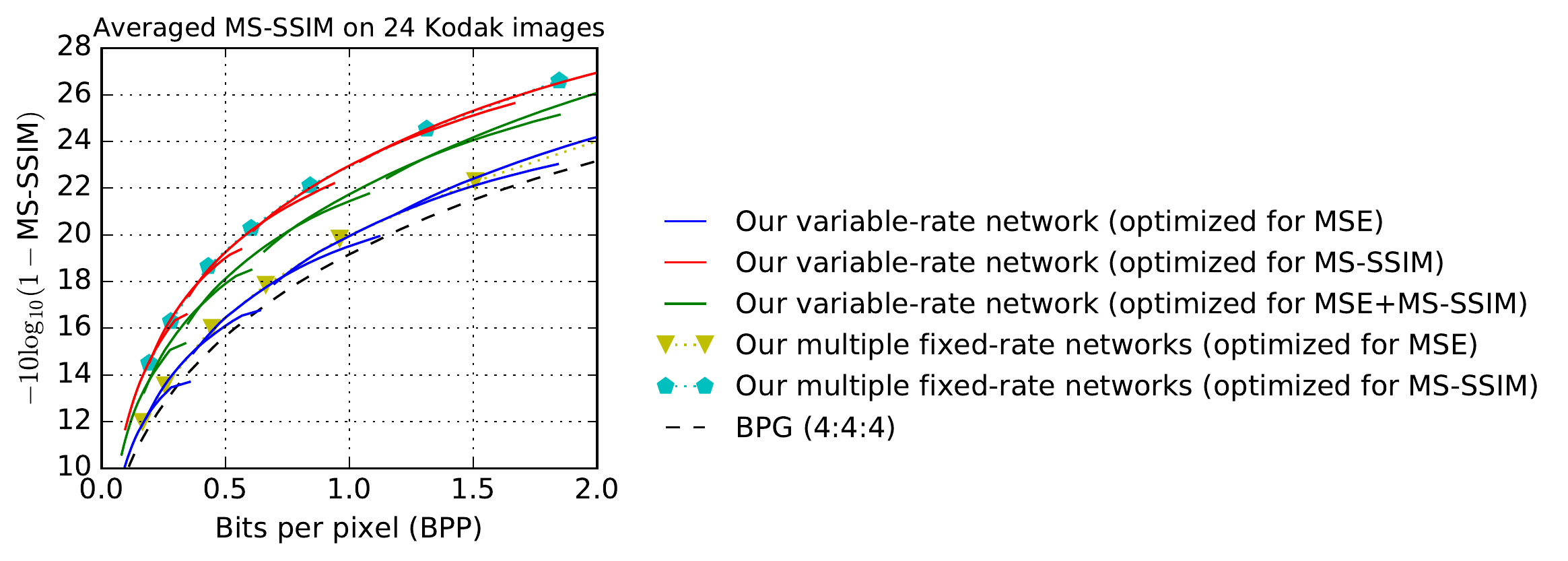}
\caption{PSNR and MS-SSIM comparison on 24 Kodak images for our variable-rate and fixed-rate networks when they are optimized for MSE, MS-SSIM and combined MSE+MS-SSIM losses, respectively. In particular, we note that our variable-rate network optimized for MSE outperforms BPG in both PNSR and MS-SSIM measures.\label{sec:exp:res:fig:02}}
\vspace{-.7em}
\end{figure*}

We pre-trained a conditional model that can be conditioned on $5$ different values of the Lagrange multiplier in $\Lambda$ for fixed bin size~$\Delta=1$, where
\begin{equation} \label{sec:exp:setup:eq:01}
\Lambda=\{10^{-1.5},10^{-2.0},10^{-2.5},10^{-3.0},10^{-3.5}\}.
\end{equation}
In pre-training, we used the MSE loss.
Then, we re-trained the model for mixed bin sizes; the quantization bin size~$\Delta$ is selected randomly from $\Delta=2^b$, where $b$ is drawn uniformly between $-1$ and $1$ so that we cover $\Delta\in[0.5,2]$. In the re-training with mixed bin sizes, we used one of MSE, MS-SSIM and combined MSE+MS-SSIM losses (see Figure~\ref{sec:exp:res:fig:02}). We used the same training datasets and the same training procedure for pre-training and re-training.
We also trained multiple fixed-rate models for fixed $\lambda\in\Lambda$ and fixed $\Delta=1$ for comparison.

\begin{figure*}[t!]
\centering
\includegraphics[height=.299\textwidth]{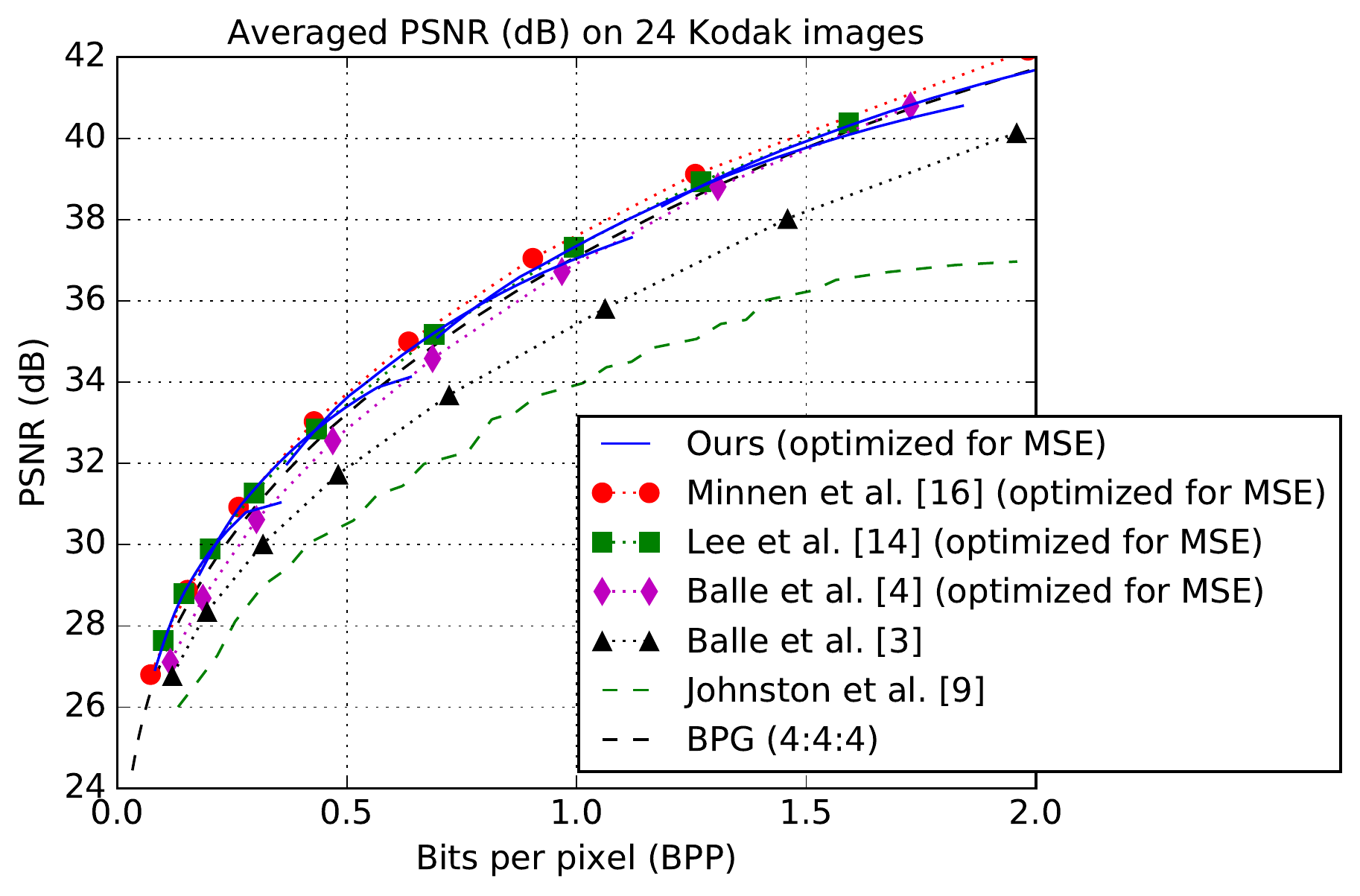}
\includegraphics[height=.299\textwidth]{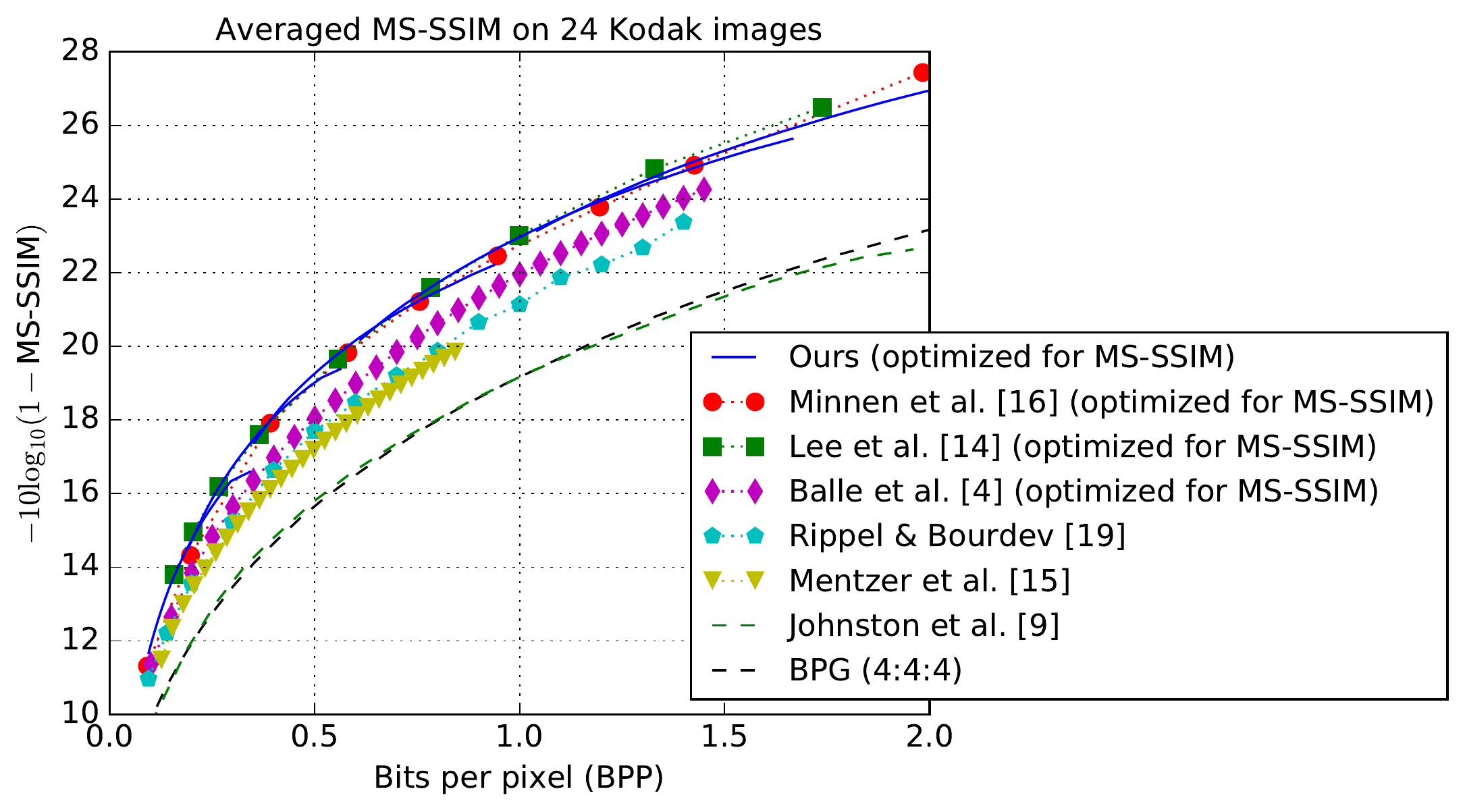}
\caption{PSNR and MS-SSIM comparison to the state-of-the-art image compression models on 24 Kodak images. As in Figure~\ref{sec:var-rate:mixedbin:fig:01}(a), we plotted $5$ curves from our variable-rate model for $5$ Lagrange multiplier values in $\Lambda$ of \eqref{sec:exp:setup:eq:01} and $\Delta\in[0.5,2]$.\label{sec:exp:res:fig:01}}
\vspace{-.7em}
\end{figure*}

\textbf{Experimental results}. We compare the performance of our variable-rate model to the state-of-the-art learned image compression models from~\cite{rippel2017real,mentzer2018conditional,balle2018variational,minnen2018joint,johnston2018improved,lee2019context} and the classical state-of-the-art variable-rate image compression codec, BPG~\cite{bellard2014bpg}, on the Kodak image set~\cite{kodak1993kodak}. Some of the previous models were optimized for MSE, and some of them were optimized for a perceptual measure, MS-SSIM. Thus, we compare both measures separately in Figure~\ref{sec:exp:res:fig:01}. In particular, we included the results for the RNN-based variable-rate compression model in \cite{johnston2018improved}, which were obtained from \cite{balle2018variational}. All the previous works in Figure~\ref{sec:exp:res:fig:01}, except \cite{johnston2018improved}, trained multiple networks to get the multiple points in their R-D curves.

For our variable-rate model, we plotted $5$ curves of the same blue color for PSNR and MS-SSIM, respectively, in Figure~\ref{sec:exp:res:fig:01}. Each curve corresponds to one of $5$ Lagrange multiplier values in \eqref{sec:exp:setup:eq:01}. For each $\lambda\in\Lambda$, we varied the quantization bin size~$\Delta$ in $[0.5,2]$ to get each curve. Our variable-rate model outperforms BPG in both PSNR and MS-SSIM measures. It also performs comparable overall and better in some cases than the state-of-the-art learned image compression models~\cite{minnen2018joint,lee2019context} that outperform BPG by deploying multiple networks trained for varying rates.

Our model shows superior performance over the RNN-based variable-rate model in \cite{johnston2018improved}. The RNN-based model requires multiple encoding/decoding iterations at high rates, implying the complexity increases as more iterations are needed to achieve better quality. In contrast, our model uses single iteration, i.e., the encoding/decoding complexity is fixed, for any rates. Moreover, our model can produce any point in the R-D curve with infinitely fine resolution by tuning the continuous rate-adaptive parameter, the quantization bin size~$\Delta$. However, the RNN-based model can produce only finite points in the R-D curve, depending on how many bits it encodes in each recurrent stage.

In Figure~\ref{sec:exp:res:fig:02}, we compare our variable-rate networks optimized for MSE, MS-SSIM and combined MSE+MS-SSIM losses, respectively. We also plotted the results from our fixed-rate networks trained for fixed $\lambda$ and $\Delta$. Observe that our variable-rate network performs very near to the ones individually optimized for fixed $\lambda$ and $\Delta$. Here, we emphasize that our variable-rate network optimized for MSE performs better than BPG in both PNSR and MS-SSIM measures.

\setlength{\tabcolsep}{0.1em}
\begin{figure*}[t!]
\centering
{\scriptsize
\begin{tabular}{ccccccccccc}
Ground truth &~~& $\lambda=10^{-3.5},\Delta=1.0$ &~~& $\lambda=10^{-2.5},\Delta=0.7$ &~~& $\lambda=10^{-2.5},\Delta=1.0$ &~~& $\lambda=10^{-2.5},\Delta=1.5$ &~~& $\lambda=10^{-1.5},\Delta=1.0$ \\
\includegraphics[width=.138\textwidth]{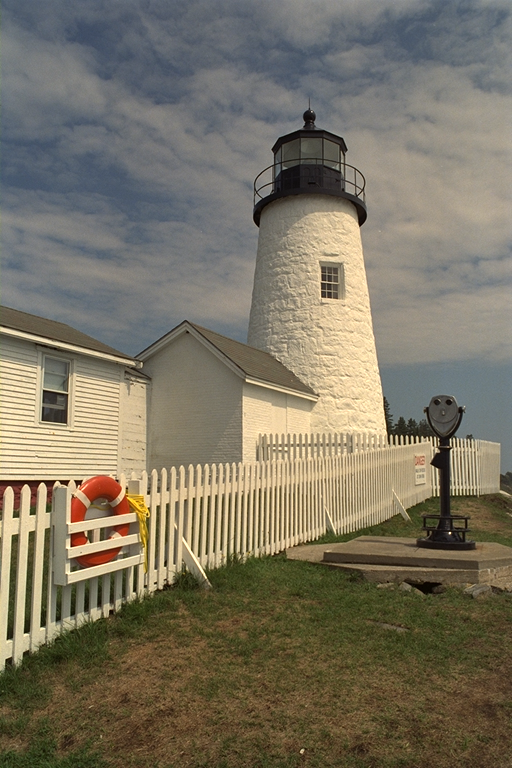}&&
\includegraphics[width=.138\textwidth]{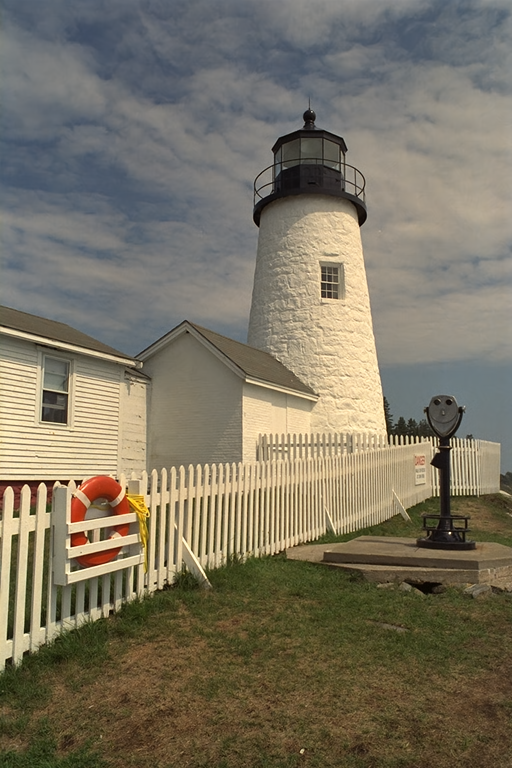}&&
\includegraphics[width=.138\textwidth]{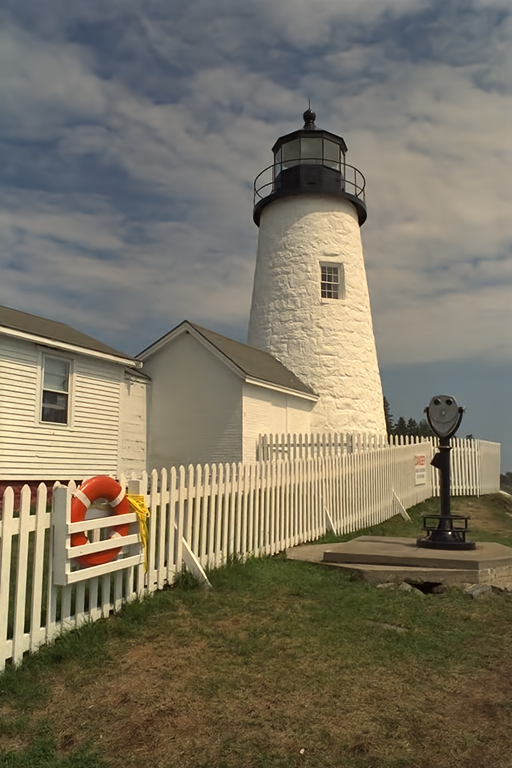}&&
\includegraphics[width=.138\textwidth]{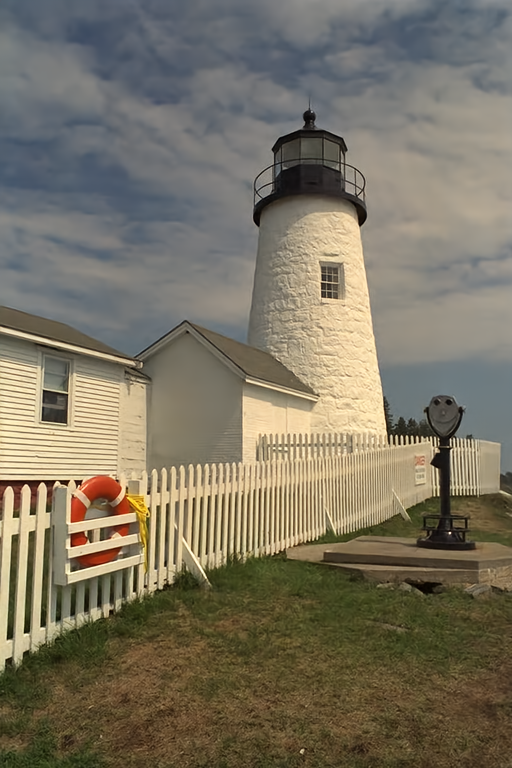}&&
\includegraphics[width=.138\textwidth]{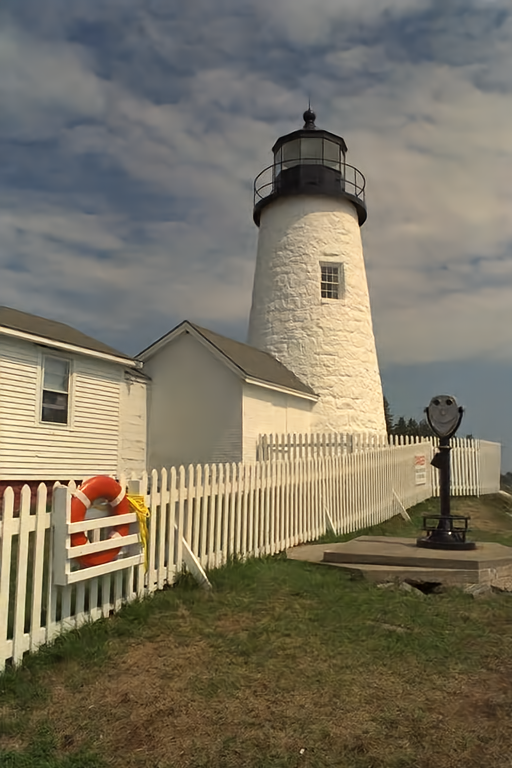}&&
\includegraphics[width=.138\textwidth]{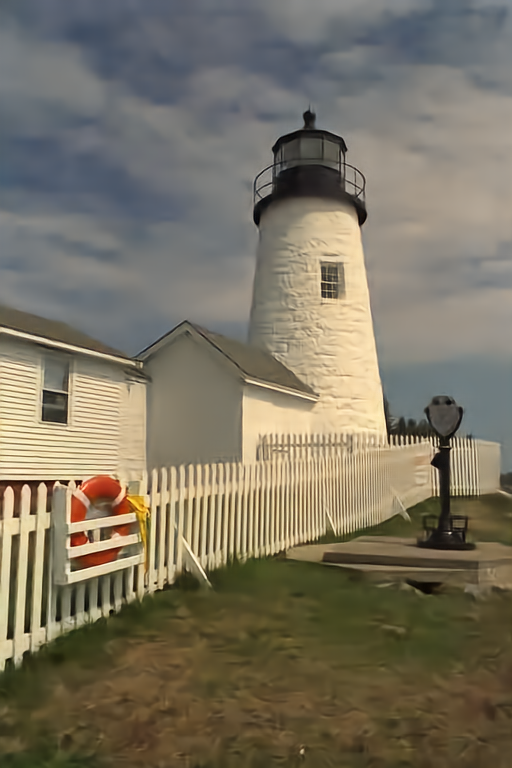}\\
\midrule
\shortstack[c]{\includegraphics[width=0.14\textwidth]{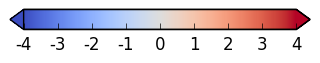}\\~\\~\\Latent representation\\$\bz$\\~}&&
\includegraphics[width=.138\textwidth]{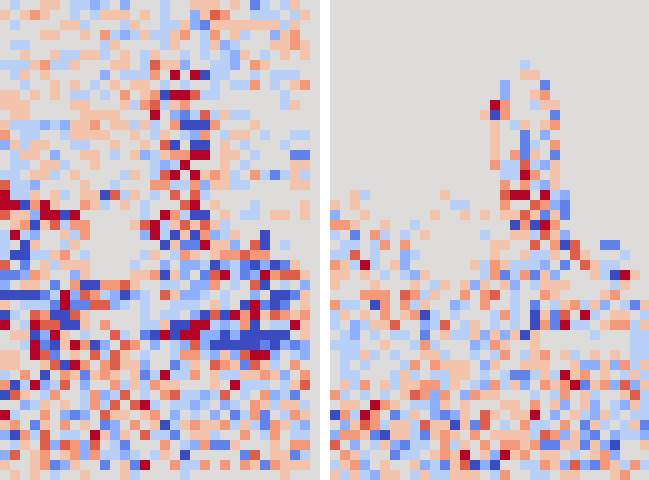}&&
\includegraphics[width=.138\textwidth]{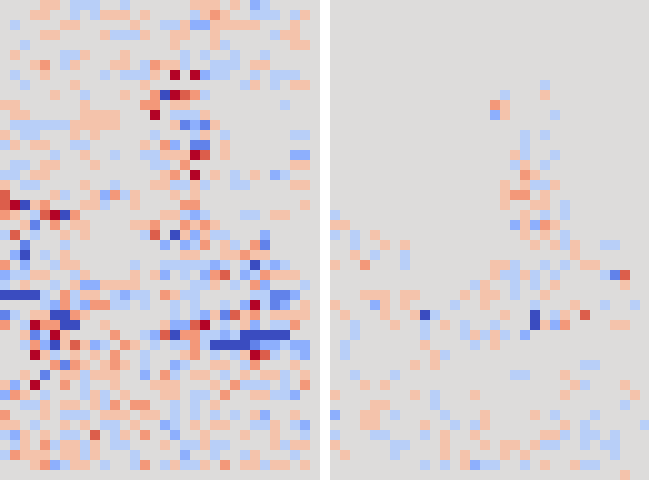}&&
\includegraphics[width=.138\textwidth]{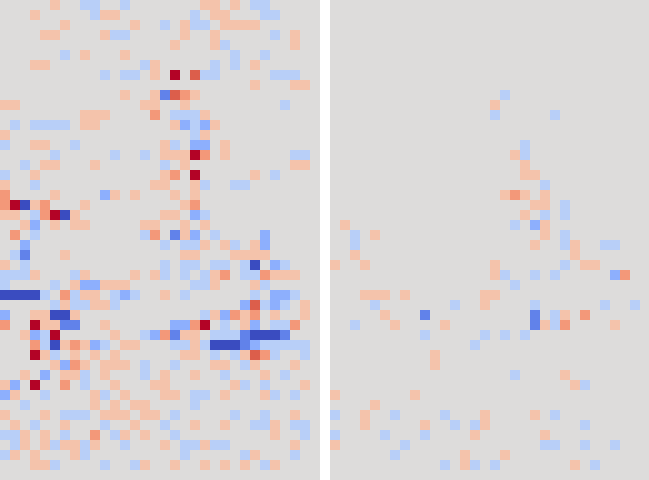}&&
\includegraphics[width=.138\textwidth]{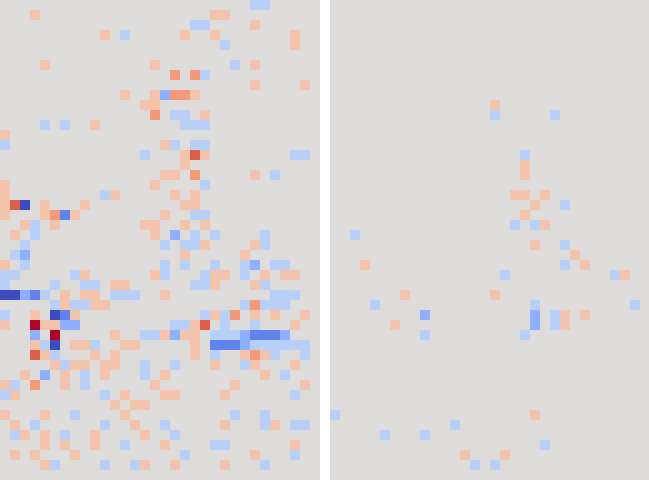}&&
\includegraphics[width=.138\textwidth]{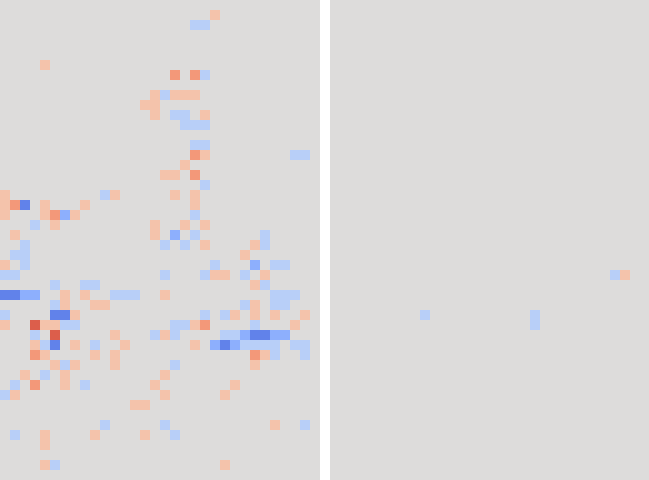}\\
\shortstack[c]{Latent representation\\$\bw$\\~}&&
\includegraphics[width=.138\textwidth]{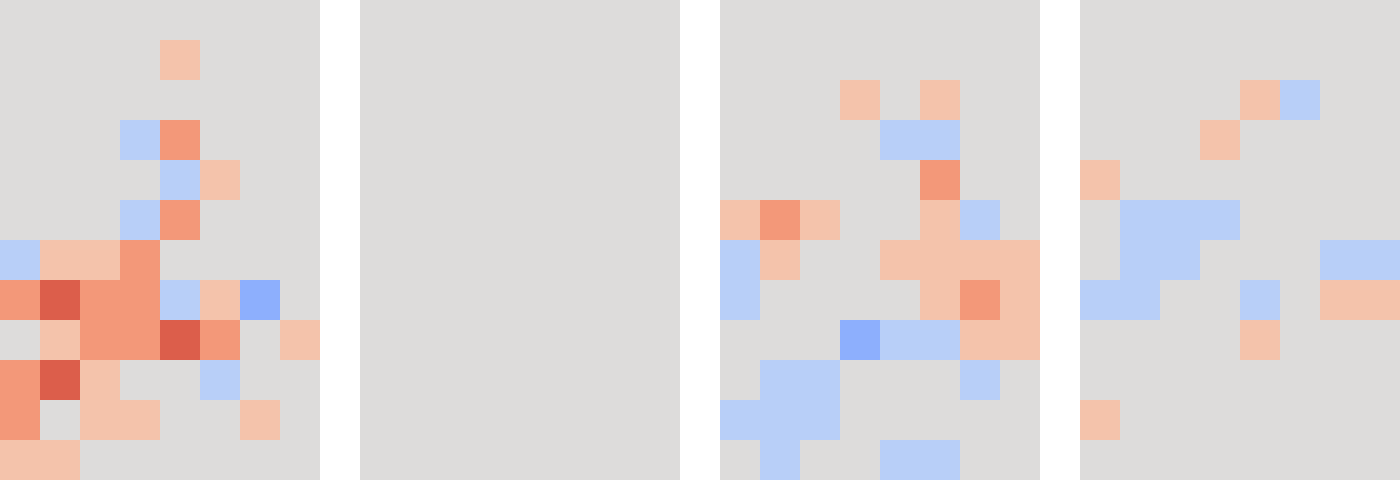}&&
\includegraphics[width=.138\textwidth]{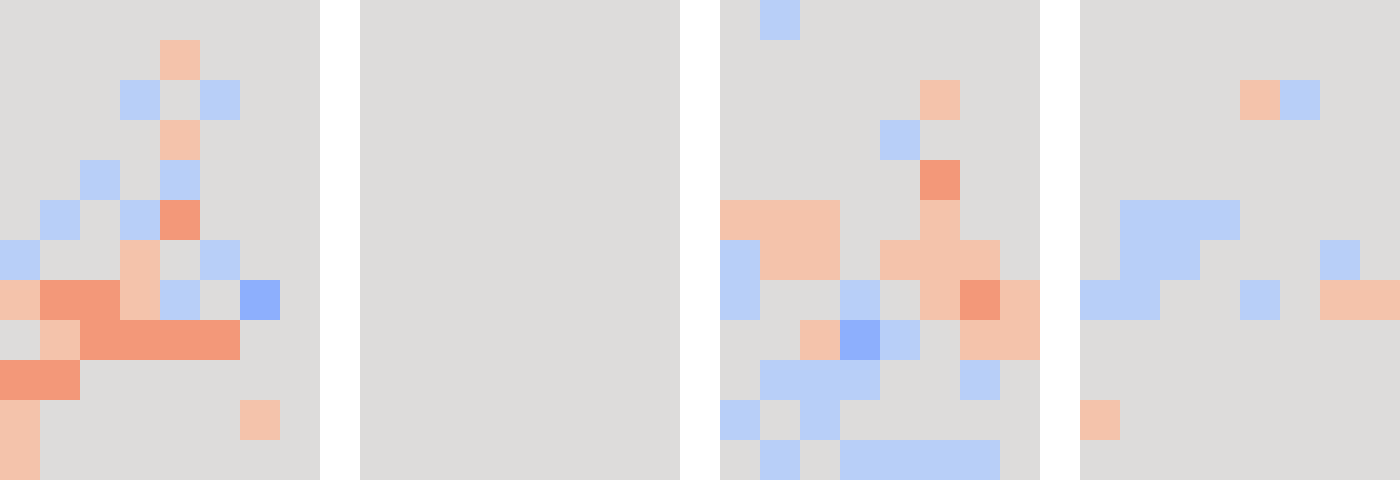}&&
\includegraphics[width=.138\textwidth]{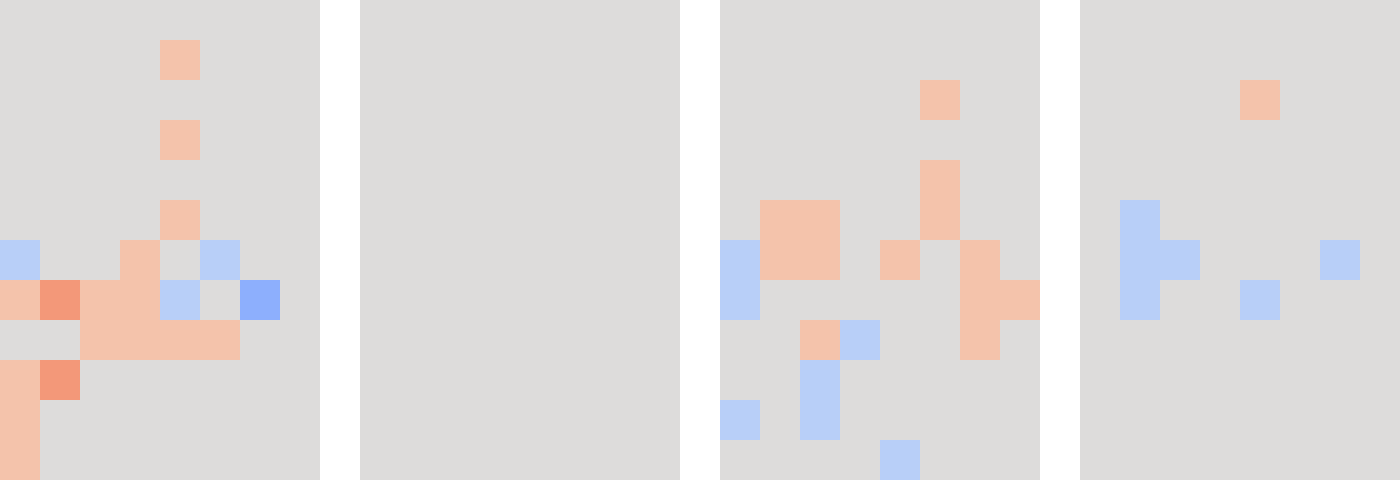}&&
\includegraphics[width=.138\textwidth]{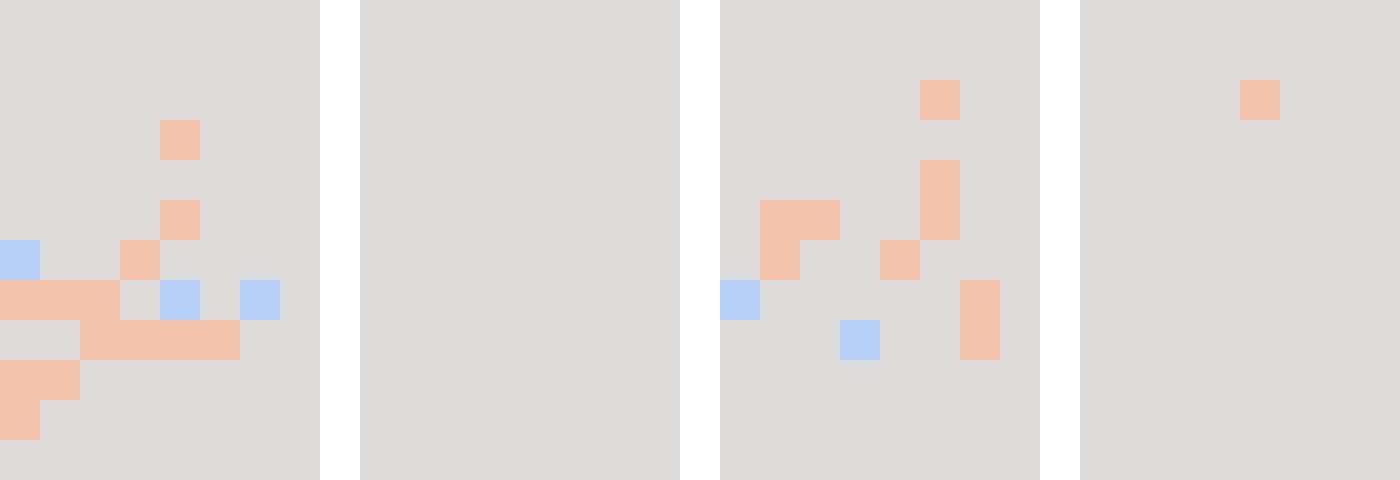}&&
\includegraphics[width=.138\textwidth]{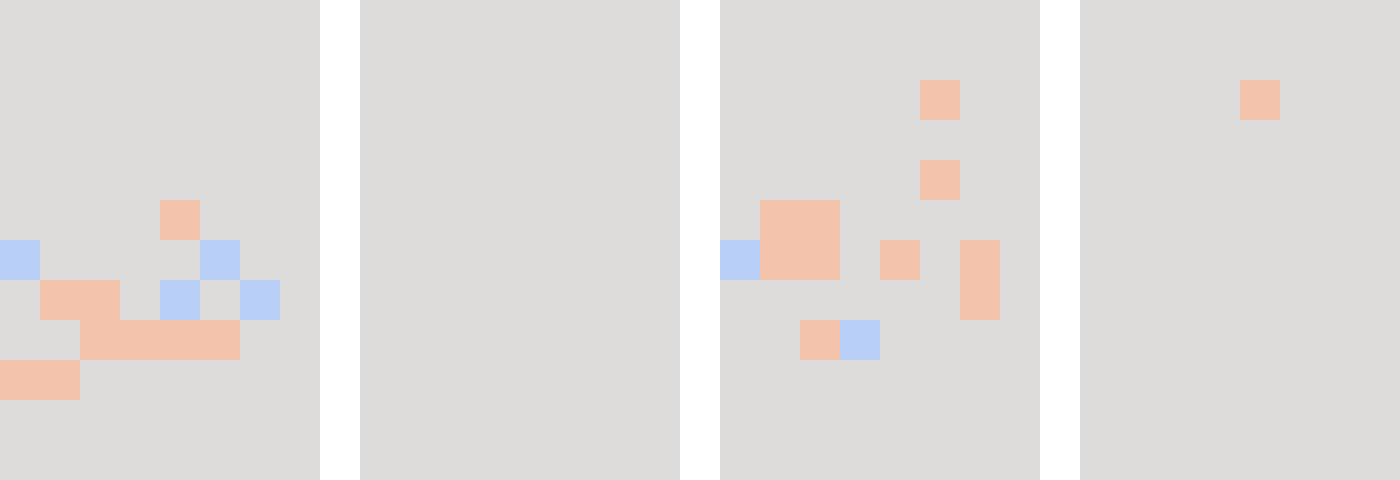}\\
\midrule
\shortstack[c]{\includegraphics[width=0.14\textwidth]{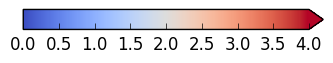}\\~\\~\\\# bits assigned to $\bz$\\in arithmetic coding\\~}&&
\includegraphics[width=.138\textwidth]{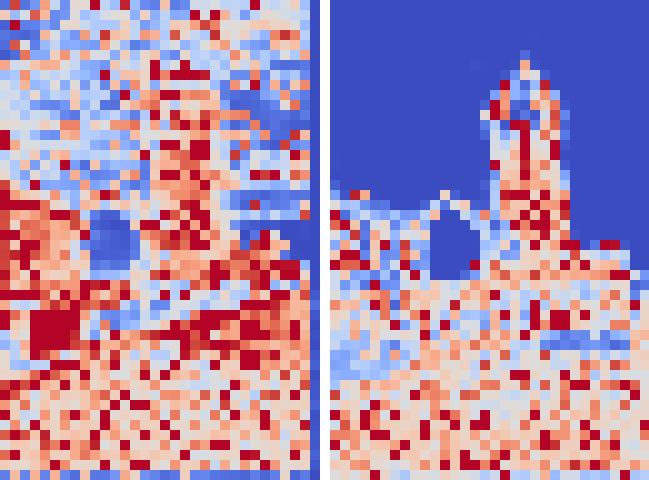}&&
\includegraphics[width=.138\textwidth]{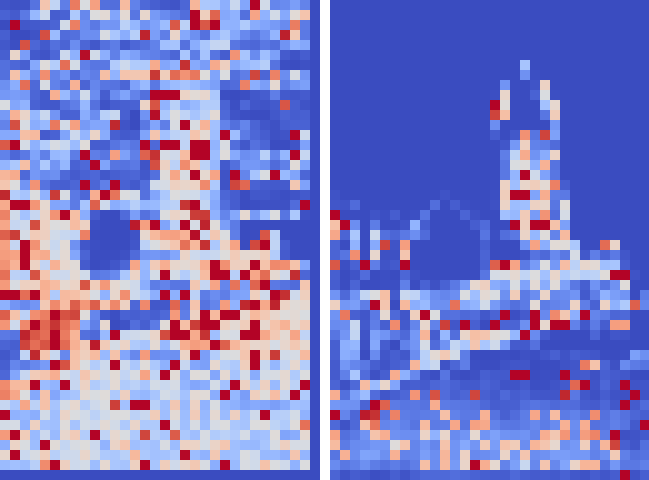}&&
\includegraphics[width=.138\textwidth]{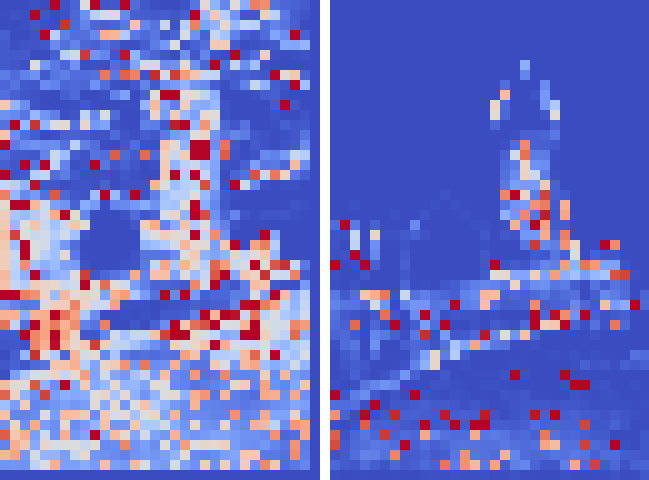}&&
\includegraphics[width=.138\textwidth]{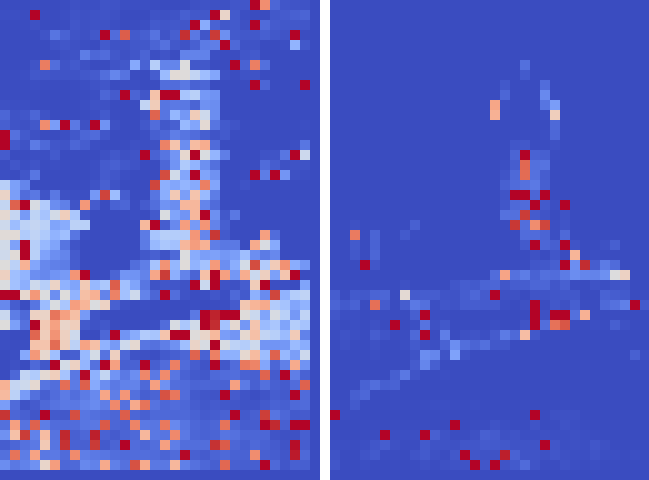}&&
\includegraphics[width=.138\textwidth]{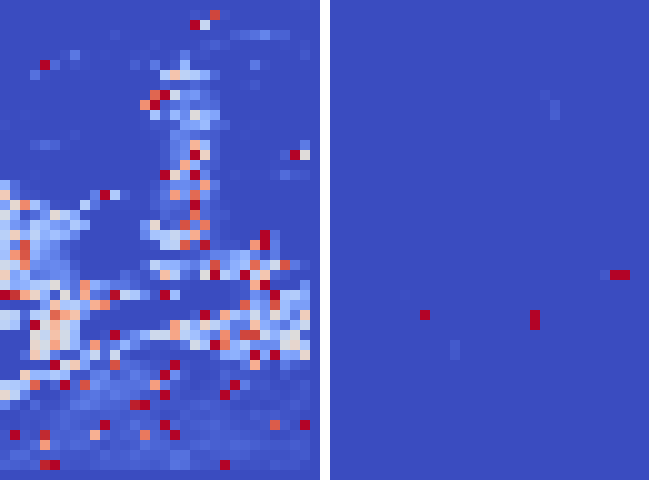}\\
\shortstack[c]{\# bits assigned to $\bw$\\in arithmetic coding\\~}&&
\includegraphics[width=.138\textwidth]{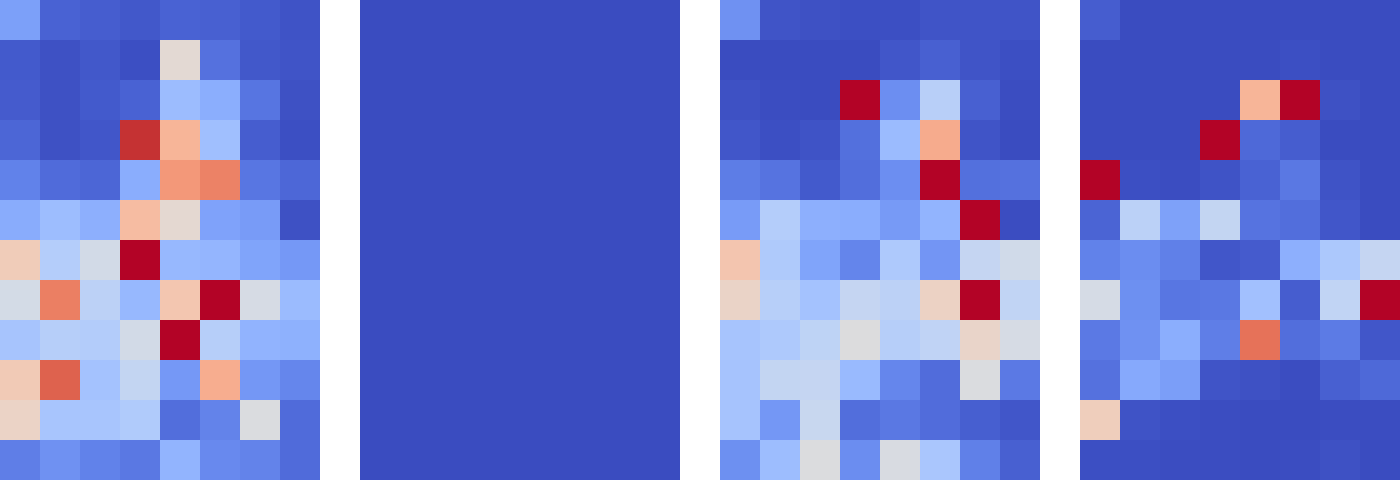}&&
\includegraphics[width=.138\textwidth]{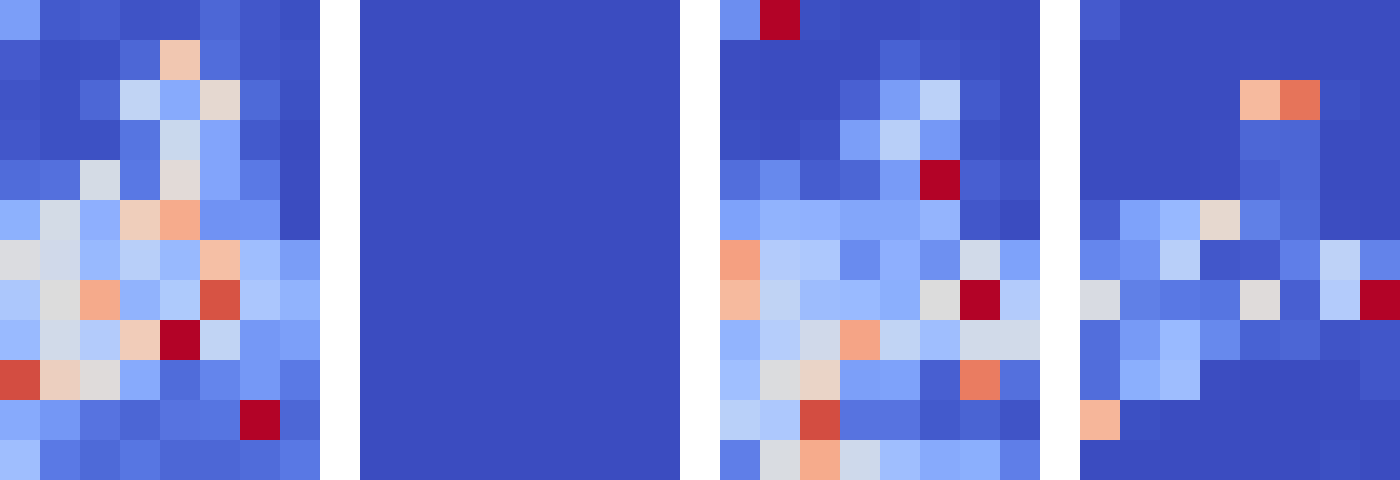}&&
\includegraphics[width=.138\textwidth]{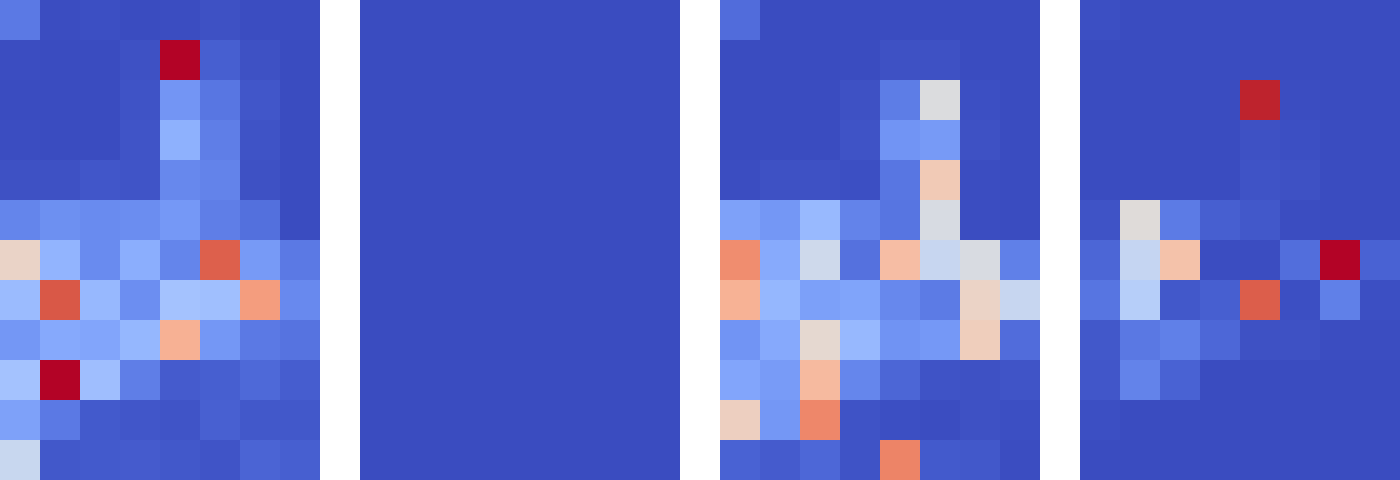}&&
\includegraphics[width=.138\textwidth]{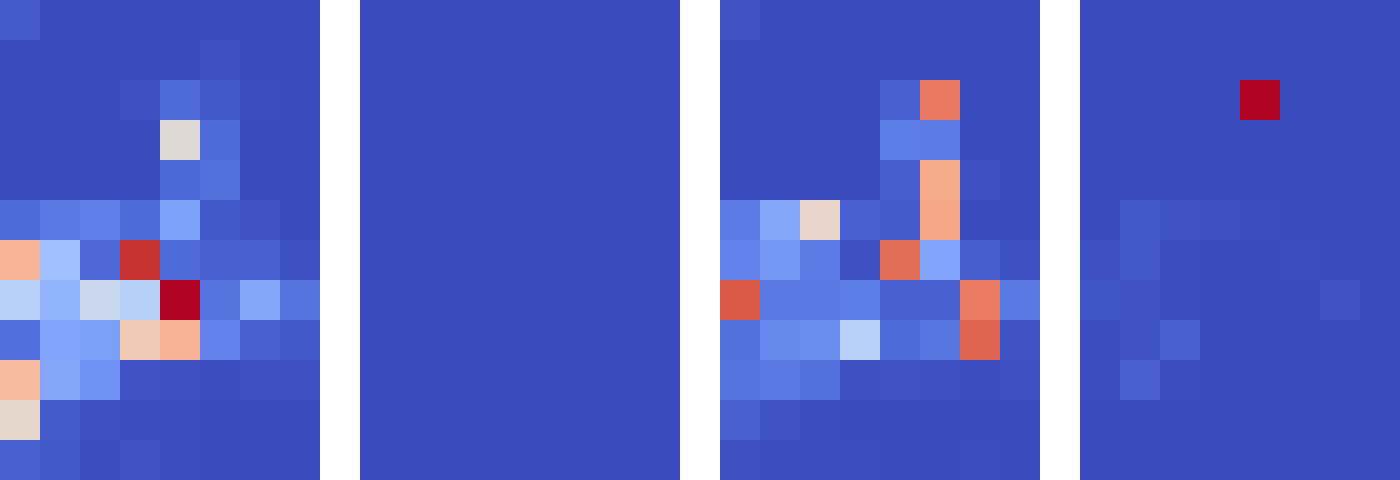}&&
\includegraphics[width=.138\textwidth]{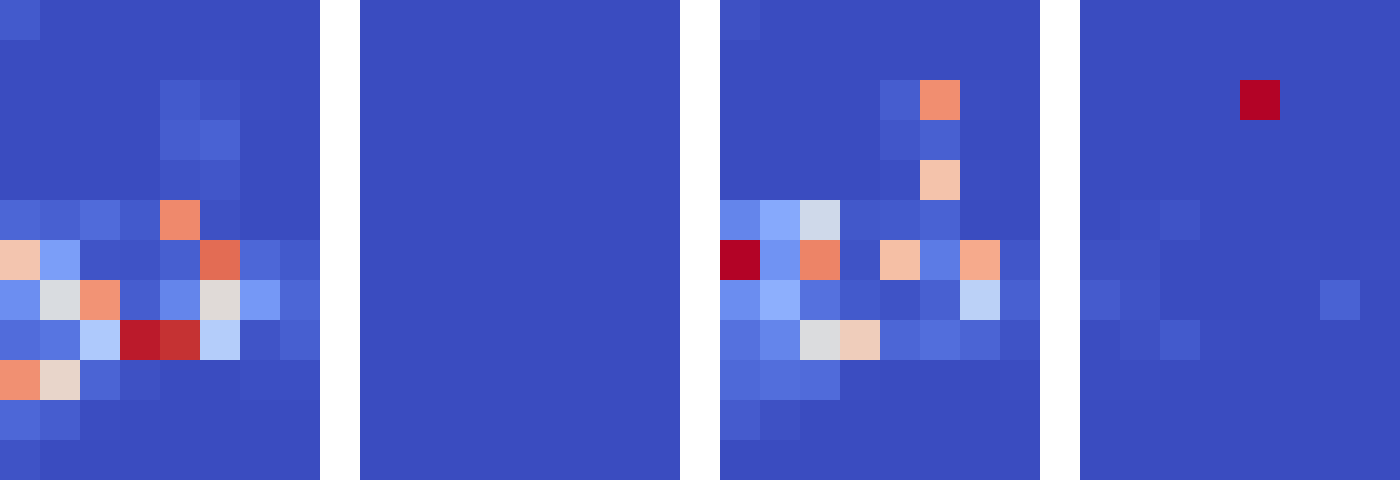}\\
\midrule
Bits per pixel (BPP) &&  1.8027 &&  0.8086 &&  0.6006 &&  0.4132 &&  0.1326\\
PSNR (dB)            && 41.3656 && 36.2535 && 34.8283 && 33.1478 && 29.2833\\
MS-SSIM              &&  0.9951 &&  0.9863 &&  0.9819 &&  0.9737 &&  0.9249\\
\bottomrule
\end{tabular}
}
\caption{Our variable-rate image compression outputs for different values of $\lambda$ and $\Delta$. We also depicted the value and the number of bits assigned to each element of latent representations~$\bz$ and $\bw$ in arithmetic coding, respectively.
\label{sec:exp:res:fig:03}}
\vspace{-.7em}
\end{figure*}


Figure~\ref{sec:exp:res:fig:03} shows the compressed images generated from our variable-rate model to assess their visual quality. We also depicted the number of bits (implicitly) used to represent each element of $\bz$ and $\bw$ in arithmetic coding, which are $-\log_2(\Delta q_\theta(z_i|z_{<i},\bw))$ and $-\log_2(\Delta q_\theta(w_i|w_{<i}))$, respectively, in \eqref{sec:refined:eq:01}--\eqref{sec:refined:eq:03}. We randomly selected two and four channels from $\bz$ and $\bw$, respectively, and showed the code length for each latent representation value in the figure. As we change conditioning parameters~$\lambda$ and $\Delta$, we can adapt the arithmetic code length that determines the rate of the latent representation. Observe that the smaller the values of $\lambda$ and/or $\Delta$, the resulting latent representation requires more bits in arithmetic coding and the rate increases, as expected.

\section{Conclusion} \label{sec:conclusion}

This paper proposed a variable-rate image compression framework with a conditional autoencoder. Unlike the previous learned image compression methods that train multiple networks to cover various rates, we train and deploy one variable-rate model that provides two knobs to control the rate, i.e., the Lagrange multiplier and the quantization bin size, which are given as input to the conditional autoencoder model. Our experimental results showed that the proposed scheme provides better performance than the classical image compression codecs such as JPEG2000 and BPG. Our method also showed comparable and sometimes better performance than the recent learned image compression methods that outperform BPG but need multiple networks trained for different compression rates. We finally note that the proposed conditional neural network can be adopted in deep learning not only for image compression but also in general to solve any optimization problem that can be formulated with the method of Lagrange multipliers.

{\small
\bibliographystyle{ieee_fullname}
\bibliography{manuscript}
}

%% file: manuscript_final_suppl.tex
%
%

\onecolumn
\appendixpage
\setcounter{section}{0}
\renewcommand*{\thesection}{\Alph{section}}
\renewcommand*{\theHsection}{app.\the\value{section}}

\section{Comparison of our refined probabilistic model to \cite{minnen2018joint}} \label{app:02}
\noindent
The major difference from \cite{minnen2018joint} is the conditioning part of $\lambda,\Delta$. Furthermore, there are some differences from \cite{minnen2018joint} in the probabilistic model, which we highlight in Table~\ref{sec:refined:tbl:01} with red color.
\setlength{\tabcolsep}{.5em}
\begin{table}[h!]
\centering
\caption{Comparison of our refined probabilistic model to \cite{minnen2018joint}.\label{sec:refined:tbl:01}}
{\small
\begin{tabular}{lll}
\toprule
Probability                 & Modeling in \cite{minnen2018joint} & Modeling in ours \\
\midrule
$p_\phi(\bw,\bz|\bx)$       & $p_\phi(\bz|\bx)p_\phi(\bw|\bz)$ & $p_\phi(\bz|\bx,{\color{red}\lambda,\Delta})p_\phi(\bw|\bz,{\color{red}\bx,\lambda,\Delta})$ \\
$q_\theta(\bx|\bw,\bz)$     & $\delta(\bx-g_\theta(\bz))$ & $\delta(\bx-g_\theta(\bz,{\color{red}\bw,\lambda}))$ \\
$q_\theta(\bz|\bw)$         & $\prod_{i}q_\theta(z_i|z_{<i},\bw)$ & $\prod_{i}q_\theta(z_i|z_{<i},\bw,{\color{red}\lambda,\Delta})$ \\
$q_\theta(\bw)$             & $\prod_{i}q_\theta(w_i)$ & $\prod_{i}q_\theta(w_i{\color{red}|w_{<i},\lambda,\Delta})$ \\
\bottomrule
\end{tabular}
}
\end{table}

\section{More example images} \label{app:03}
\begin{figure*}
\centering
{\scriptsize
\begin{tabular}{ccc}
\includegraphics[width=0.333\textwidth]{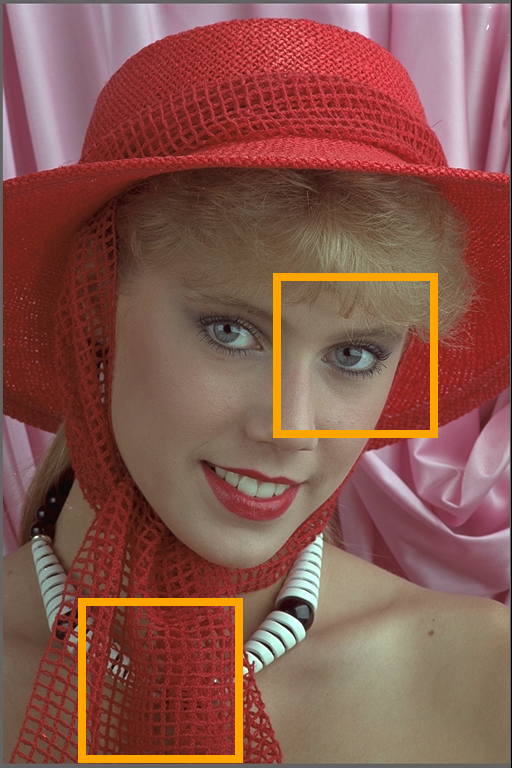} & \includegraphics[width=0.333\textwidth]{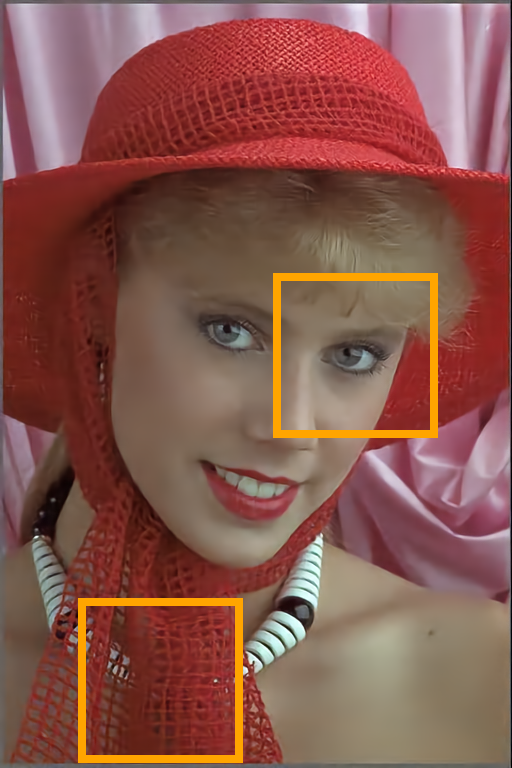} & \includegraphics[width=0.333\textwidth]{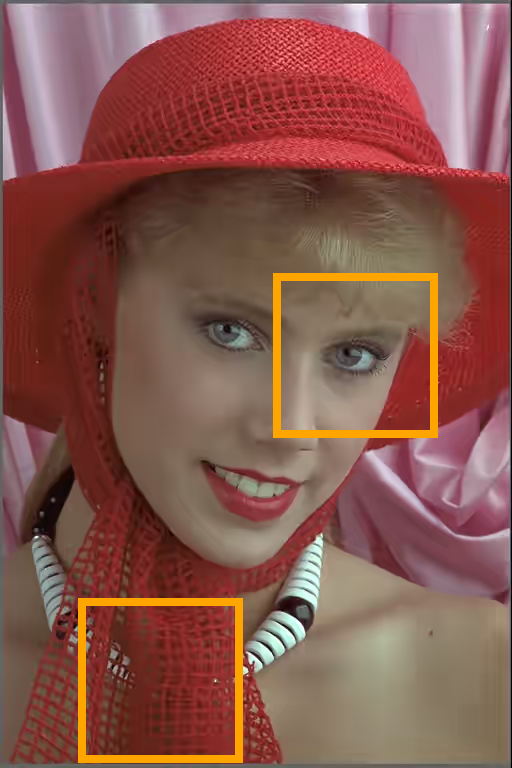} \\
\shortstack[c]{Ground truth\\~} & \shortstack[c]{Ours\\BPP: 0.2078, PSNR: 32.4296 (dB), MS-SSIM: 0.9543} & \shortstack[c]{BPG (4:4:4)\\BPP: 0.2078, PSNR: 32.0406 (dB), MS-SSIM: 0.9488} \\
\midrule
& \includegraphics[width=0.333\textwidth]{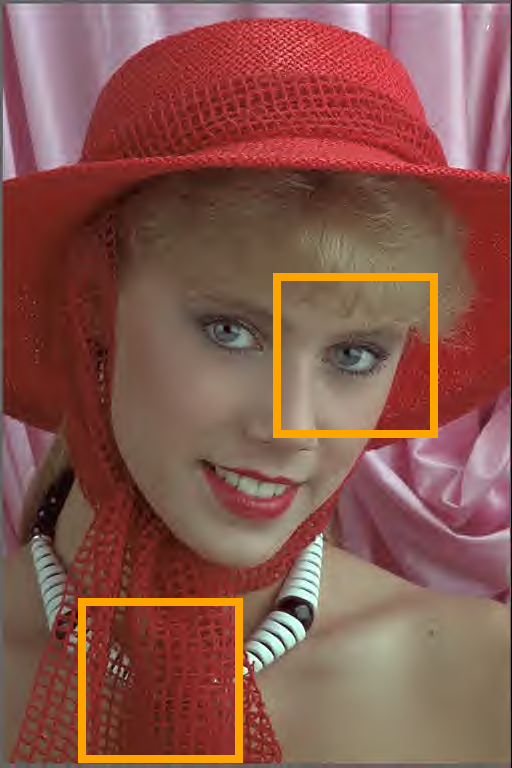} & \includegraphics[width=0.333\textwidth]{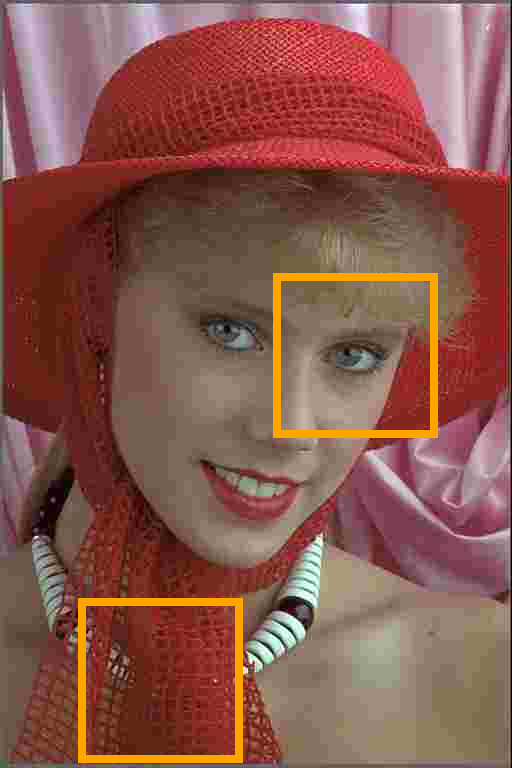} \\
& \shortstack[c]{JPEG2000\\BPP: 0.2092, PSNR: 30.9488 (dB), MS-SSIM: 0.9342} & \shortstack[c]{JPEG\\BPP: 0.2098, PSNR: 28.1758 (dB), MS-SSIM: 0.8777} \\
\bottomrule
\end{tabular}
}
\caption{PSNR, MS-SSIM, and visual quality comparison of our variable-rate deep image compression method and classical image compression algorithms (BPG, JPEG2000, and JPEG) for the Kodak image 04. Our method outperforms the state-of-the-art BPG codec in both PSNR and MS-SSIM measures. We put orange boxes to highlight the visual differences.\label{sec:app:fig:03}}
\vspace{-.7em}
\end{figure*}

\begin{figure*}
\centering
{\scriptsize
\begin{tabular}{ccccc}
Ground truth & Ours & BPG (4:4:4) & JPEG2000 & JPEG \\
\includegraphics[width=0.2\textwidth]{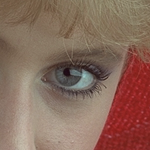} &
\includegraphics[width=0.2\textwidth]{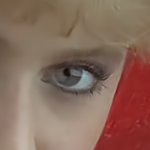} &
\includegraphics[width=0.2\textwidth]{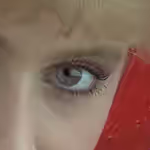} &
\includegraphics[width=0.2\textwidth]{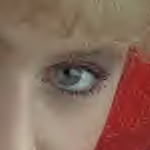}  &
\includegraphics[width=0.2\textwidth]{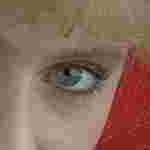} \\
\includegraphics[width=0.2\textwidth]{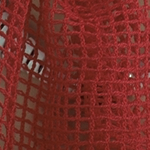} &
\includegraphics[width=0.2\textwidth]{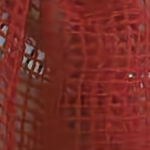} &
\includegraphics[width=0.2\textwidth]{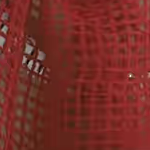} &
\includegraphics[width=0.2\textwidth]{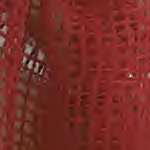}  &
\includegraphics[width=0.2\textwidth]{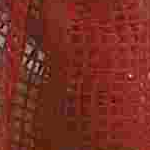} \\
\toprule
Bits per pixel (BPP) &  0.2078 &  0.2078 &  0.2092 &  0.2098 \\
PSNR (dB)            & 32.4296 & 32.0406 & 30.9488 & 28.1758 \\
MS-SSIM              &  0.9543 &  0.9488 &  0.9342 &  0.8777 \\
\bottomrule
\end{tabular}
}
\caption{Visual quality comparison of our variable-rate deep image compression method and classical image compression algorithms (BPG, JPEG2000, and JPEG) for the Kodak image 04 in the orange-boxed areas of Figure~\ref{sec:app:fig:03}.\label{sec:app:fig:03a}}
\vspace{-.7em}
\end{figure*}

\begin{figure*}
\centering
{\scriptsize
\begin{tabular}{cc}
\includegraphics[width=0.5\textwidth]{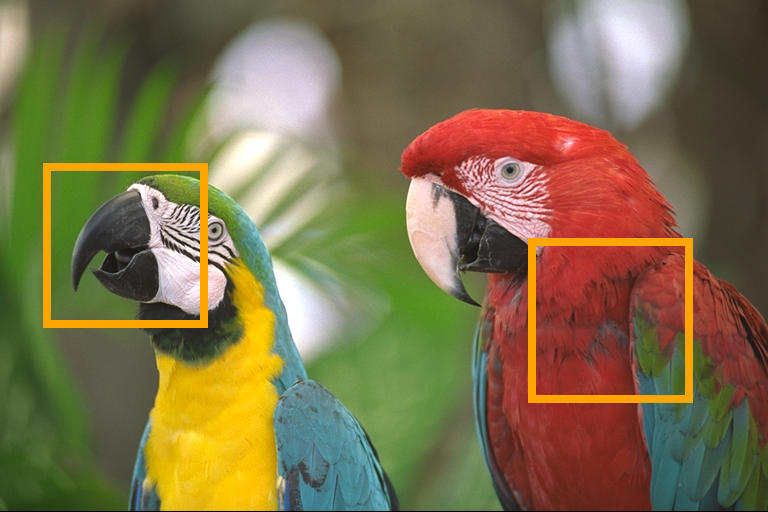} \\
Ground truth \\
\midrule
\includegraphics[width=0.5\textwidth]{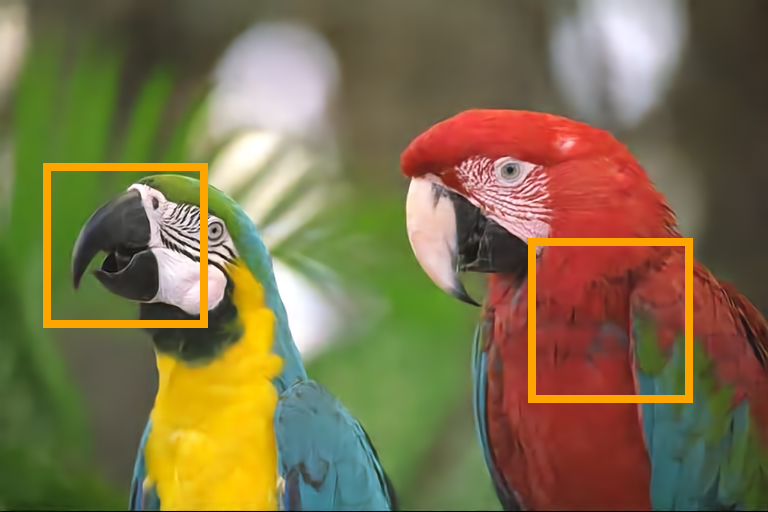} & \includegraphics[width=0.5\textwidth]{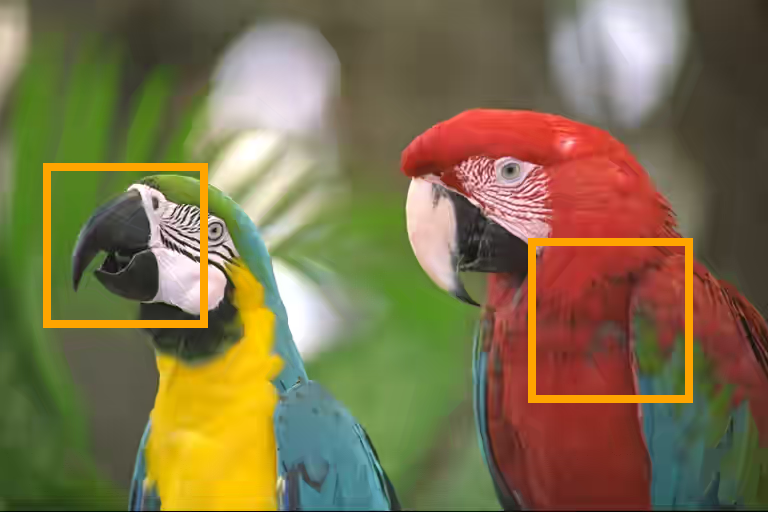} \\
\shortstack[c]{Ours\\BPP: 0.1289, PSNR: 34.4543 (dB), MS-SSIM: 0.9695} & \shortstack[c]{BPG (4:4:4)\\BPP: 0.1289, PSNR: 33.3546 (dB), MS-SSIM: 0.9593} \\
\midrule
\includegraphics[width=0.5\textwidth]{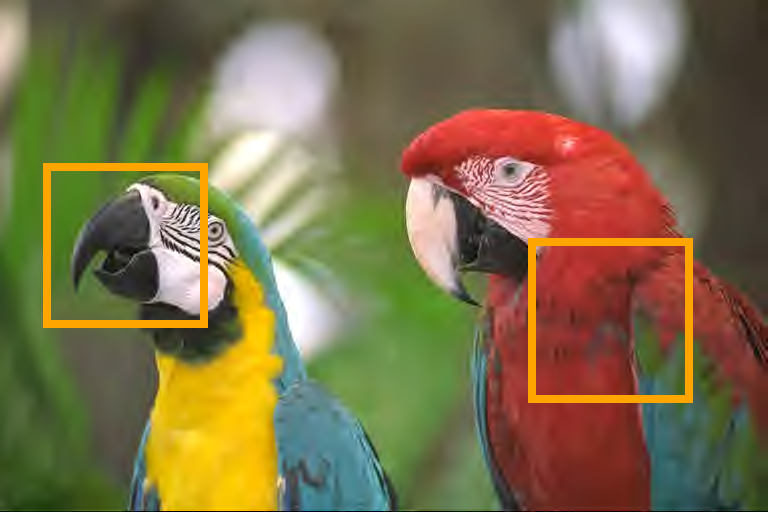} & \includegraphics[width=0.5\textwidth]{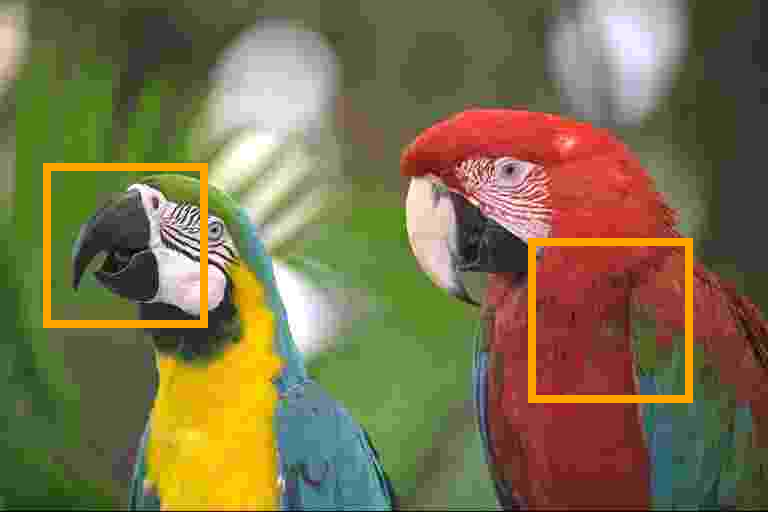} \\
\shortstack[c]{JPEG2000\\BPP: 0.1298, PSNR: 31.8927 (dB), MS-SSIM: 0.9482} & \shortstack[c]{JPEG\\BPP: 0.1299, PSNR: 27.1270 (dB), MS-SSIM: 0.8404} \\
\bottomrule
\end{tabular}
}
\caption{PSNR, MS-SSIM, and visual quality comparison of our variable-rate deep image compression method and classical image compression algorithms (BPG, JPEG2000, and JPEG) for the Kodak image 23. Our method outperforms the state-of-the-art BPG codec in both PSNR and MS-SSIM measures. We put orange boxes to highlight the visual differences.\label{sec:app:fig:07}}
\vspace{-.7em}
\end{figure*}

\begin{figure*}
\centering
{\scriptsize
\begin{tabular}{ccccc}
Ground truth & Ours & BPG (4:4:4) & JPEG2000 & JPEG \\
\includegraphics[width=0.2\textwidth]{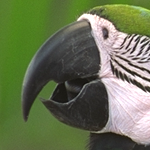} &
\includegraphics[width=0.2\textwidth]{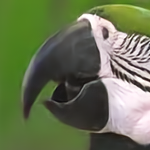} &
\includegraphics[width=0.2\textwidth]{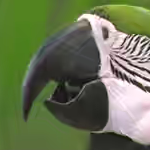} &
\includegraphics[width=0.2\textwidth]{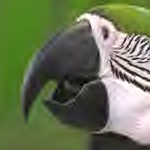}  &
\includegraphics[width=0.2\textwidth]{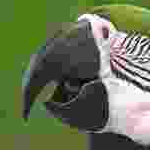} \\
\includegraphics[width=0.2\textwidth]{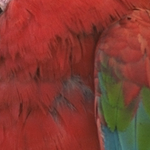} &
\includegraphics[width=0.2\textwidth]{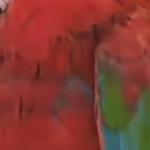} &
\includegraphics[width=0.2\textwidth]{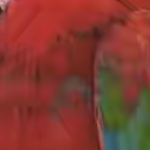} &
\includegraphics[width=0.2\textwidth]{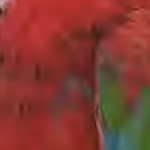}  &
\includegraphics[width=0.2\textwidth]{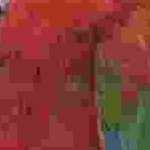} \\
\toprule
Bits per pixel (BPP) &  0.1289 &  0.1289 &  0.1298 &  0.1299 \\
PSNR (dB)            & 34.4543 & 33.3546 & 31.8927 & 27.1270 \\
MS-SSIM              &  0.9695 &  0.9593 &  0.9482 &  0.8404 \\
\bottomrule
\end{tabular}
}
\caption{Visual quality comparison of our variable-rate deep image compression method and classical image compression algorithms (BPG, JPEG2000, and JPEG) for the Kodak image 23 in the orange-boxed areas of Figure~\ref{sec:app:fig:07}.\label{sec:app:fig:07a}}
\vspace{-.7em}
\end{figure*}

\noindent
As supplementary materials, we provide more example images produced by our variable-rate image compression network that is optimized for the MSE loss. We compare our method to the classical image compression codecs, i.e., JPEG, JPEG2000, and BPG. We adapt and match the compression rate of our variable-rate network to the rate of BPG by adjusting the Lagrange multiplier~$\lambda$ and the quantization bin size~$\Delta$. All the examples show that our method outperforms the state-of-the-art BPG codec in both PSNR and MS-SSIM measures at the same bits per pixel (BPP). Visually, our method provides better quality with less artifacts than the classical image compression codecs. We put orange boxes to highlight the visual differences in Figure~\ref{sec:app:fig:03} and Figure~\ref{sec:app:fig:07}, and the orange-boxed areas are magnified in Figure~\ref{sec:app:fig:03a} and Figure~\ref{sec:app:fig:07a}, respectively. 

\setlength{\tabcolsep}{0.1em}

\setlength{\tabcolsep}{0.1em}

\setlength{\tabcolsep}{0.1em}

\setlength{\tabcolsep}{0.1em}
